\documentclass[aps,prl,reprint]{revtex4-2}

\usepackage{graphicx}
\usepackage{amsmath,amssymb} 
\usepackage{physics}
\graphicspath{figures/}
\usepackage{dcolumn}
\usepackage{amssymb}
\usepackage{amsmath}
\usepackage{fancyhdr}
\usepackage{bm}
\usepackage{times}
\usepackage[colorlinks,linkcolor=blue,anchorcolor=blue,citecolor=blue,urlcolor=blue]{hyperref}

\begin{document}

\title{Photon blockade via three-body interactions: toward high-purity and bright single-photon sources}
\author{Sheng Zhao}
\affiliation{Ministry of Education Key Laboratory for Nonequilibrium Synthesis and Modulation of Condensed Matter, Shaanxi Province Key Laboratory of Quantum Information and Quantum Optoelectronic Devices, School of Physics, Xi'an Jiaotong University, Xi'an 710049, China}
\author{Peng-Bo Li}
\email{lipengbo@mail.xjtu.edu.cn}
\affiliation{Ministry of Education Key Laboratory for Nonequilibrium Synthesis and Modulation of Condensed Matter, Shaanxi Province Key Laboratory of Quantum Information and Quantum Optoelectronic Devices, School of Physics, Xi'an Jiaotong University, Xi'an 710049, China}

\date{\today}
\begin{abstract}
Photon blockade is vital for single-photon generation, but current schemes with conventional and unconventional photon blockade face critical limitations like the purity-brightness trade-off, hindering the generation of high-performance single-photons. To overcome these limitations, we introduce a fundamentally new photon blockade mechanism by utilizing three-body interactions between a single photonic mode and two qubits. This kind of interaction intrinsically cuts off the excitation path to the two-photon state, resulting in a perfect photon blockade effect. The mechanism operates across a broad parameter range, free from the constraints of strong coupling or weak driving. Remarkably, it \textit{breaks the purity-brightness trade-off}, enabling the simultaneous achievement of extreme purity and high brightness, both significantly outperforming previous mechanisms. Furthermore, this approach demonstrates robustness against thermal noise and avoids unwanted oscillations in the time-delayed correlation function. This work provides a path for generating high-purity, high-brightness, and robust single-photon sources, a key resource for quantum technologies.
\end{abstract}
\maketitle
\textit{Introduction}---Photon blockade (PB) is a profound quantum phenomenon wherein the absorption of a single photon strictly forbids the subsequent excitation of others~\cite{PhysRevLett.109.193602, PhysRevLett.129.043601, SanchezMunoz2014, Lingenfelter2021,PhysRevLett.107.223601, PhysRevB.87.235319,Shen2015,PhysRevA.87.023822,PhysRevLett.121.153601,PhysRevA.104.053718,PhysRevA.100.053857,PhysRevA.100.043831,PhysRevLett.125.073601,chakram2022multimode,PhysRevLett.123.013602,PhysRevLett.134.013602,dong2025collectiveenhancementphotonblockade}. It serves as a crucial physical mechanism for generating high-quality single-photon sources~\cite{ Lounis2005, Aharonovich2016, Wang2019, Senellart2017, Reimer2019, RevModPhys.87.347,PhysRevX.9.021016,PhysRevLett.122.186804}, which are indispensable fundamental resources for optical quantum computing~\cite{Knill2001, Zhong2020}, quantum communication~\cite{Kimble2008, Yin2020}, and quantum sensing~\cite{Degen2017, Polino2020}. The canonical mechanism to realize PB, widely known as conventional photon blockade (CPB), originates from the strong anharmonicity of the energy spectrum~\cite{Imamoglu1997, Rabl2011,PhysRevA.82.032101, Liao2013, Wang2015, Zou:20,PhysRevLett.118.133604,PhysRevA.102.033713}. This approach fundamentally relies on the strong nonlinear coupling of an optical mode to atoms~\cite{Birnbaum2005}, quantum dots~\cite{Faraon2008}, and superconducting qubits~\cite{Lang2011, PhysRevLett.107.053602, Fink2017}, necessitating that the coupling strength significantly surpasses the dissipation rate of the optical mode. Severely limited by this demanding coupling condition, CPB remains largely out of reach in many systems that exhibit only weak nonlinearity.

To circumvent this limitation and realize PB in weakly coupled systems, unconventional photon blockade (UPB) was proposed, which relies on destructive quantum interference between distinct excitation pathways~\cite{Liew2010, Bamba2011, PhysRevA.88.033836, Flayac2017, Wang2021, Lemonde2014, Gerace2014, Xu2014,PhysRevA.96.053827, Hou2019, Li:19, PhysRevA.98.023856,PhysRevLett.122.243602,PhysRevLett.125.197402,10.1063/5.0289640,PhysRevA.108.023716,PhysRevA.92.023838}. Unfortunately, the requisite exact interference imposes highly demanding parameter matching~\cite{Majumdar2012, Ferretti2013,Wang:23}, and the time-delayed second-order correlation function exhibits a drastically narrow anti-bunching time window, making its experimental observation severely limited by finite detector time resolution~\cite{PhysRevLett.121.043601, Vaneph2018}. Moreover, UPB yields a vanishingly small average photon number, which inevitably results in extremely low source brightness. Beyond their respective drawbacks, CPB, UPB, and even their simultaneous occurrence~\cite{PhysRevLett.134.183601} share a critical limitation: any attempt to enhance the brightness by increasing the driving strength causes the second-order correlation function to rise simultaneously, thereby severely degrading the single-photon purity. Consequently, existing PB mechanisms remain trapped in an inescapable trade-off between single-photon purity and brightness. Therefore, exploring novel PB mechanisms to break this trade-off and realize high-purity, high-brightness single-photon sources, particularly in the weak-coupling regime unconstrained by coupling strength, is highly desirable.

In this work, we propose and analyze a previously unidentified
photon blockade mechanism for achieving near-perfect PB based on a three-body interaction involving a single photonic mode and two qubits, with the high-frequency qubit being coherently driven. Specifically, under this Hamiltonian, the excitation path from the ground state to the single-photon state is allowed, but the transition path from the single-photon state to the two-photon state is cut off, leading to a strong blockade effect. We further employ the second-order correlation function $g^{(2)}(0)$ and the average photon number $N$ to quantify the performance of this scheme. Compared to previous CPB and UPB mechanisms, our scheme offers three critical advantages: 
(i) It operates efficiently across a broad parameter space, free from the strict constraints of strong coupling or weak driving. The mechanism avoids the reliance on both the spectral anharmonicity required by CPB and the exact destructive interference that imposes demanding parameter matching in UPB. It thus exhibits remarkable robustness against parameter variations, maintaining strong antibunching regardless of the specific coupling or driving strengths.
(ii) It successfully breaks the purity-brightness trade-off, achieving significantly higher brightness and purity than previous methods. Notably, in the weak-coupling regime, $g^{(2)}(0)$ remains nearly constant over a broad
range of driving strengths while $N$ increases continuously. The onset of the $g^{(2)}(0)$ rise and the saturation of $N$ occur in close proximity, enabling the nearly simultaneous realization of the minimum $g^{(2)}(0)$ and the maximum $N$. Consequently, our scheme yields an emission rate far surpassing that of UPB (limited by a vanishingly small $N$) and CPB (restricted by a small photon decay rate). 
(iii) It maintains strong resistance to thermal noise and strictly avoids unwanted oscillations in the time-delayed correlation function, thereby ensuring robust and high-quality single-photon emission.
Our scheme can be implemented in a variety of platforms, ranging from cavity QED to microwave photonic setups.

\textit{The model and mechanisms}---We investigate a hybrid quantum system consisting of a single photonic mode and two qubits, described by a three-body interaction model. The free Hamiltonian of the system is given by
$\hat{H}_0 = \omega_a \hat{a}^\dagger \hat{a} + \omega_1 \hat{\sigma}_1^+ \hat{\sigma}_1^- + \omega_2 \hat{\sigma}_2^+ \hat{\sigma}_2^-$, where $\hat{a}$ ($\hat{a}^\dagger$) is the annihilation (creation) operator of the photon with frequency $\omega_a$. 
For the $j$-th qubit ($j = 1,2$), the operators $\hat{\sigma}_j^-$ ($\hat{\sigma}_j^+$) denote the lowering (raising) operator, and $\omega_j$ is its transition frequency. The photon interacts with the two qubits via a three-body coupling, described by the interaction Hamiltonian
\begin{equation}
\hat{H}_{\text{thr}} = J \left( \hat{a}^\dagger \hat{\sigma}_1^+ \hat{\sigma}_2^- + \hat{a} \hat{\sigma}_1^- \hat{\sigma}_2^+ \right),
\label{eq1}
\end{equation}
where $J$ is the coupling strength of the three-body interaction. Under the resonance condition $\omega_2=\omega_1+\omega_a$, the interaction induces the transition between $|n-1, g_1, e_2\rangle $ and $ |n, e_1, g_2\rangle$, where $|n\rangle$ is the photon Fock state and $|g_j\rangle$ ($|e_j\rangle$) denotes the ground (excited) state of the $j$-th qubit. This describes the excitation of the photon and the first qubit upon deexcitation of the second qubit and the inverse process. The interaction Hamiltonian can be diagonalized in each subspace spanned by $\{|n-1, g_1, e_2\rangle, |n, e_1, g_2\rangle\}$. The resulting dressed states are $
|\Psi_n^\pm\rangle =( |n-1, g_1, e_2\rangle \pm |n, e_1, g_2\rangle )/\sqrt{2}$, which
satisfy $\hat{H}_{\text{thr}} |\Psi_n^\pm\rangle=E_{n}^\pm|\Psi_n^\pm\rangle$ with eigenenergies $E_{n}^\pm = \pm J\sqrt{n}$. The three-body interaction thus induces level splittings of $2J\sqrt{n}$.

To realize and manipulate PB, a drive field is applied to the second qubit. The total Hamiltonian of the system reads $\hat{H}_{\text{sys}} = \hat{H}_0 + \hat{H}_{\text{thr}} + \hat{H}_{d}$, where $\hat{H}_{d} = \Omega (\hat{\sigma}_2^+ e^{-i\omega_d t} + \hat{\sigma}_2^- e^{i\omega_d t})$ describes the driving term, with $\omega_d$ being the drive frequency and $\Omega$ denoting the driving strength. In the rotating frame with respect to $V = \omega_a \hat{a}^\dagger \hat{a} + (\omega_d - \omega_a) \hat{\sigma}_1^+ \hat{\sigma}_1^- + \omega_d \hat{\sigma}_2^+ \hat{\sigma}_2^-$, the total Hamiltonian becomes:
\begin{align}
\hat{H}_{\text{tot}} = &\Delta \hat{\sigma}_1^+ \hat{\sigma}_1^- + \Delta \hat{\sigma}_2^+ \hat{\sigma}_2^- 
+ J \bigl( \hat{a}^\dagger \hat{\sigma}_1^+ \hat{\sigma}_2^- + \hat{a} \hat{\sigma}_1^- \hat{\sigma}_2^+ \bigr) \nonumber \\
&+ \Omega \bigl( \hat{\sigma}_2^+ + \hat{\sigma}_2^- \bigr),
\end{align}
where $\Delta=\omega_2-\omega_d$ is the detuning of the second qubit with respect to the drive frequency.

\begin{figure}[t]
    \centering
    \includegraphics[width=0.5\textwidth]{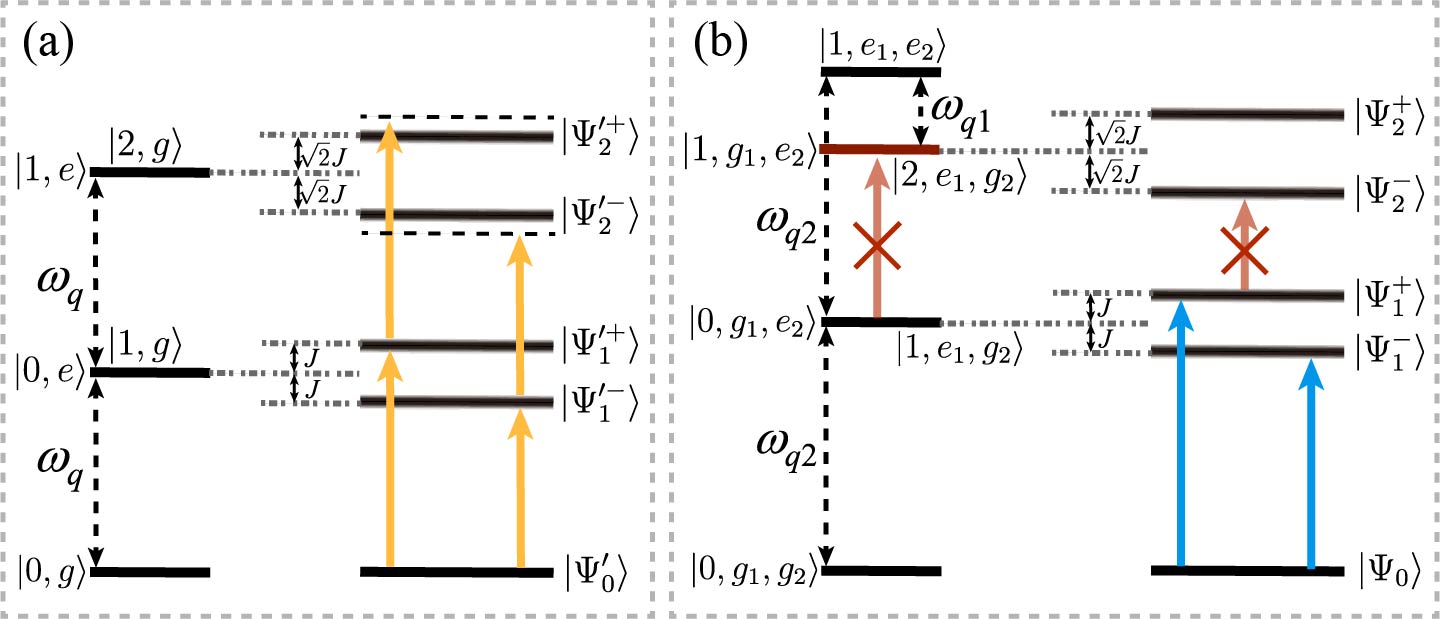}
    \caption{(a) The mechanism of CPB originated from energy-level anharmonicity. (b) Novel mechanism of PB based on three-body interaction, where no transition path exists between single-photon and two-photon states.}
    \label{FIG1}
\end{figure}

As shown in Fig.~\ref{FIG1}(a), the CPB relies on energy-level anharmonicity. Strong coupling results in an anharmonic ladder of dressed states, making the single-photon transition resonant while detuning the two-photon transition. Consequently, CPB requires the coupling strength to greatly exceed the dissipation rates to ensure sufficient anharmonicity. 
In stark contrast, our PB mechanism stems from the fact that there is no transition matrix element connecting the one-photon and two-photon subspaces under the three-body Hamiltonian. As shown in Fig.~\ref{FIG1}(b), the drive applied to the second qubit excites the system from the ground state $|0, g_1, g_2\rangle$ to $|0, g_1, e_2\rangle$. The three-body interaction then couples $|0, g_1, e_2\rangle$ and $|1, e_1, g_2\rangle$, forming the single-photon dressed state $|\Psi_1^\pm\rangle$, thereby bringing the system into the single-photon state. The drive can subsequently excite the system to $|1, e_1, e_2\rangle$. However, from this state, any transition to a two-photon state is forbidden: because the first qubit is already excited, it is impossible to generate a second photon via the deexcitation of the second qubit; i.e., the photon-number-increasing operator $\hat{a}^\dagger \hat{\sigma}_1^+ \hat{\sigma}_2^-$ acting on $|1, e_1, e_2\rangle$ yields exactly zero (In contrast, for the standard JC interaction, the transition $|1,e\rangle \to |2,g\rangle$ is allowed). Thus, under this Hamiltonian, the system can be excited from the ground state to the single-photon state $|\Psi_1^\pm\rangle$, but no transition pathway exists between the single-photon state $|\Psi_1^\pm\rangle$ and the two-photon state $|\Psi_2^\pm\rangle$. Because our mechanism does not rely on anharmonicity, it is therefore not restricted by the coupling strength.

\textit{Results of photon blockade}---For an open quantum system, the evolution of its density matrix $\rho$ is governed by the master equation
\begin{align}
\frac{\partial}{\partial t}\rho = -i[ \hat{H}_{\text{tot}}, \rho] + \kappa \mathcal{L}_{\hat{a}}[ \rho ] + \gamma \sum_{i=1,2} \mathcal{L}_{\hat{\sigma}_i^-}[ \rho ],   
\end{align}
where $\mathcal{L}_{\hat{o}}[ \rho ] = \hat{o}\rho\hat{o}^\dagger - (\hat{o}^{\dagger}\hat{o}\rho + \rho \hat{o}^{\dagger}\hat{o})/2$ is the Lindblad superoperator, $\kappa$ is the decay rate of the photonic mode, and $\gamma$ is the decay rate of each qubit. In the presence of dissipation, the population of the two-photon state does not remain strictly zero, since qubit decay relaxes the system from $|1, e_1, e_2\rangle$ to $|1, g_1, e_2\rangle$ and opens a weak pathway to two-photon states, though this effect remains small. We focus on the steady state solution $\rho_s$, which is computed numerically using the QuTiP package~\cite{johansson2012qutip}. For analytical insight, the Hilbert space is truncated to the two-photon subspace, since the excitation of higher Fock states is strongly suppressed under the total Hamiltonian, allowing the steady-state equations to be solved analytically~\cite{SM}.

\begin{figure*}[t]
    \centering
    \includegraphics[width=0.9\textwidth]{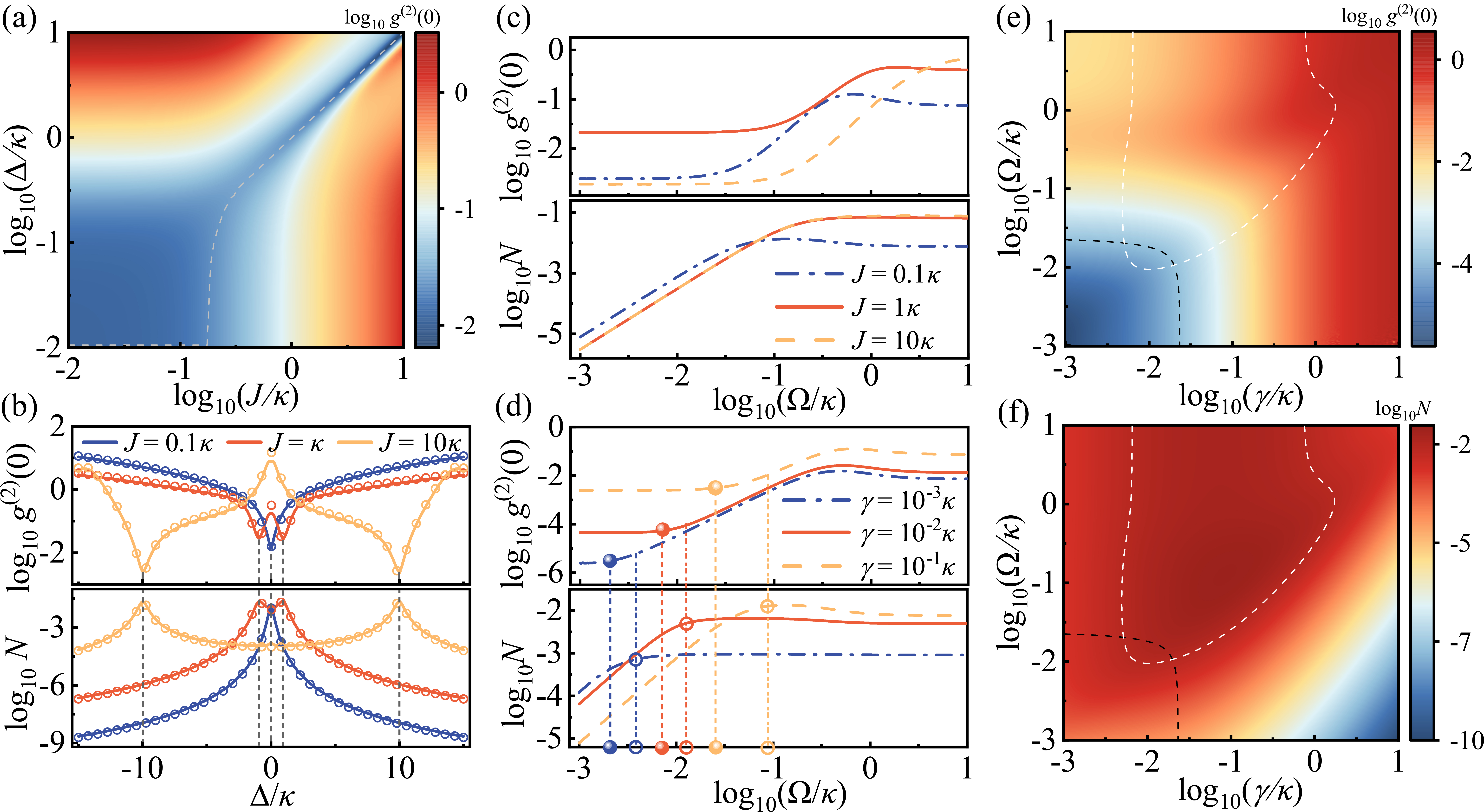}
    \caption{(a) Color map of $\log_{10} g^{(2)}(0)$ versus normalized coupling $J/\kappa$ and detuning $\Delta/\kappa$. The dashed line marks the optimal detuning trajectory, corresponding to the minimal $g^{(2)}(0)$ at each fixed $J/\kappa$. (b) Variations of $\log_{10} g^{(2)}(0)$ and $\log_{10} N$ with $\Delta/\kappa$ for different coupling strengths $J$. For (a) and (b), we take $\Omega=0.1\kappa$ and $\gamma=0.1\kappa$. (c, d) $\log_{10} g^{(2)}(0)$ and $\log_{10} N$ versus $\Omega$ for (c) different coupling strengths $J$ with $\gamma=0.1\kappa$ and (d) different qubit dissipation rates $\gamma$ with $J=0.1\kappa$. Solid and open circles denote the onset of the $g^{(2)}(0)$ rise and the saturation point of $N$, respectively. (e, f) Color maps of (e) $\log_{10} g^{(2)}(0)$ and (f) $\log_{10} N$ in the $\Omega$-$\gamma$ parameter space at $J=0.1\kappa$ and $\Delta=0$. The black and white dashed contours indicate $\log_{10} g^{(2)}(0) = -3.8$ and $\log_{10} N = -2.2$, respectively.}
    \label{FIG2}
\end{figure*}

To characterize the photon statistical properties, we employ the equal-time second-order correlation function $g^{(2)}(0) = \langle \hat{a}^\dagger \hat{a}^\dagger \hat{a} \hat{a} \rangle / \langle \hat{a}^\dagger \hat{a} \rangle^2$, with $g^{(2)}(0) \to 0$ signaling PB. Figure.~\ref{FIG2}(a) shows the numerical steady-state $g^{(2)}(0)$ as a function of $J/\kappa$ and $\Delta/\kappa$. Strong photon antibunching is observed over a broad parameter region. As expected, the PB is not restricted by the coupling strength, with $g^{(2)}(0) \approx 10^{-2}$ reached even at $J/\kappa = 10^{-2}$. The optimal detuning for PB is $\Delta/\kappa = 0$ in the weak-coupling limit, and equals the coupling strength $J$ in the strong-coupling regime, as indicated by the white dashed line in Fig~\ref{FIG2}(a). Going further, we derive the analytical solution for the equal-time second-order correlation function
\begin{equation}   
g^{(2)}(0) = \frac{2\mathcal{N}}{N^2},
\end{equation}
where $\mathcal{N}$ is the two-photon population (see Supplemental Material for details~\cite{SM}), and $N $ is the mean photon number (dominated by the single-photon population), given by 
\begin{equation}
    N\approx\frac{J^2\Omega^2}{(J^2-\Delta^2)^2+\Delta^2\kappa^2/4+J^2\kappa\gamma+\alpha\Omega^2},
\end{equation}
with $\alpha = J^2\kappa/\gamma+3\kappa^2/4$. Clearly, $N$ is maximized at $\Delta=0$ for weak coupling and $|\Delta|=J$ for strong coupling, which directly leads to the minimal $g^{(2)}(0)$, as confirmed by Fig.~\ref{FIG2}(b). So, the optimal detuning simultaneously optimizes both the single-photon purity (minimized $g^{(2)}(0)$) and the source brightness, which is proportional to the mean photon number $N$. Physically, optimal PB is achieved when the drive resonates with the single-photon dressed state to maximize the single-photon population, while two-photon excitation is suppressed by the absence of the direct excitation path. Furthermore, the analytical results (symbols) are in excellent agreement with the numerical results (solid lines) in Fig.~\ref{FIG2}(b).

Next, we investigate the influence of the driving strength $\Omega$ on the photon antibunching. As shown in Fig.~\ref{FIG2}(c), $g^{(2)}(0)$ remains nearly constant at low driving strengths, then increases gradually with increasing $\Omega$, and finally saturates at a constant value. In the strong-coupling regime ($J \geq \kappa$), the PB disappears under strong driving ($\Omega > \kappa$). Significantly, in the weak-coupling regime, $g^{(2)}(0)$ remains below $0.1$ over the entire driving range, ensuring a robust PB even under strong driving, in stark contrast to CPB and UPB, which strictly require weak driving. Furthermore, under the condition of strong PB with $g^{(2)}(0) < 0.01$, the weak-coupling regime yields a high mean photon number $N \approx 10^{-2}$, comparable to that under strong coupling, making it an advantageous working region. We thus focus subsequent analysis on the weak-coupling regime. In the weak-coupling limit with optimal detuning $\Delta=0$, the analytical expressions for $N$ and $\mathcal{N}$ can be simplified as
\begin{align}
 N\approx\frac{1}{(J^{2}+\kappa\gamma)/\Omega^{2}+\kappa/\gamma+3\kappa^{2}/4J^{2}},\nonumber\\
\mathcal {N}\approx\frac{21J^{4}\gamma^{2}[J^{2}\gamma^{2}\Omega^{4}+(1.5J^{2}+6.5\gamma^{2})\Omega^{6}]}{\kappa^{4}(\Xi+28J^{2}\kappa^{2}\gamma^{4}\Omega^{2}+\Theta\Omega^{4})},
\end{align}
with parameters $\Xi=2J^{6}\kappa\gamma^{3}+42J^{6}\gamma^{4}+J^{4}\kappa^{2}\gamma^{4}$ and $\Theta=2J^{4}\kappa^{2}+1.5J^{2}\kappa^{3}\gamma+500J^{2}\kappa\gamma^{3}+6\kappa^{3}\gamma^{3}+80\kappa^{2}\gamma^{4}$. According to the analytical solution, $N\propto\Omega^2$ at low driving strengths, then grows slowly and eventually saturates as the driving-dependent term weakens with increasing $\Omega$. The two-photon population evolves from $\mathcal{N}\propto\Omega^4$ to $\mathcal{N}\propto\Omega^6$ as the drive strengthens. Consequently, $g^{(2)}(0)$, governed by the ratio $\mathcal{N}/N^2$, remains nearly constant in the weak-driving regime (while $N$ grows rapidly as $\Omega^2$), and then scales as $g^{(2)}(0)\propto\Omega^2$ before approaching saturation. Notably, the onset of the $g^{(2)}(0)$ rise slightly precedes the saturation of $N$, indicating that their optimal operating regions are in close proximity. This asynchronous growth is in stark contrast to the inherent trade-off in CPB and UPB, where $N$ and $g^{(2)}(0)$ grow almost synchronously.

\begin{figure}[t]    
\centering    
\includegraphics[width=0.5\textwidth]{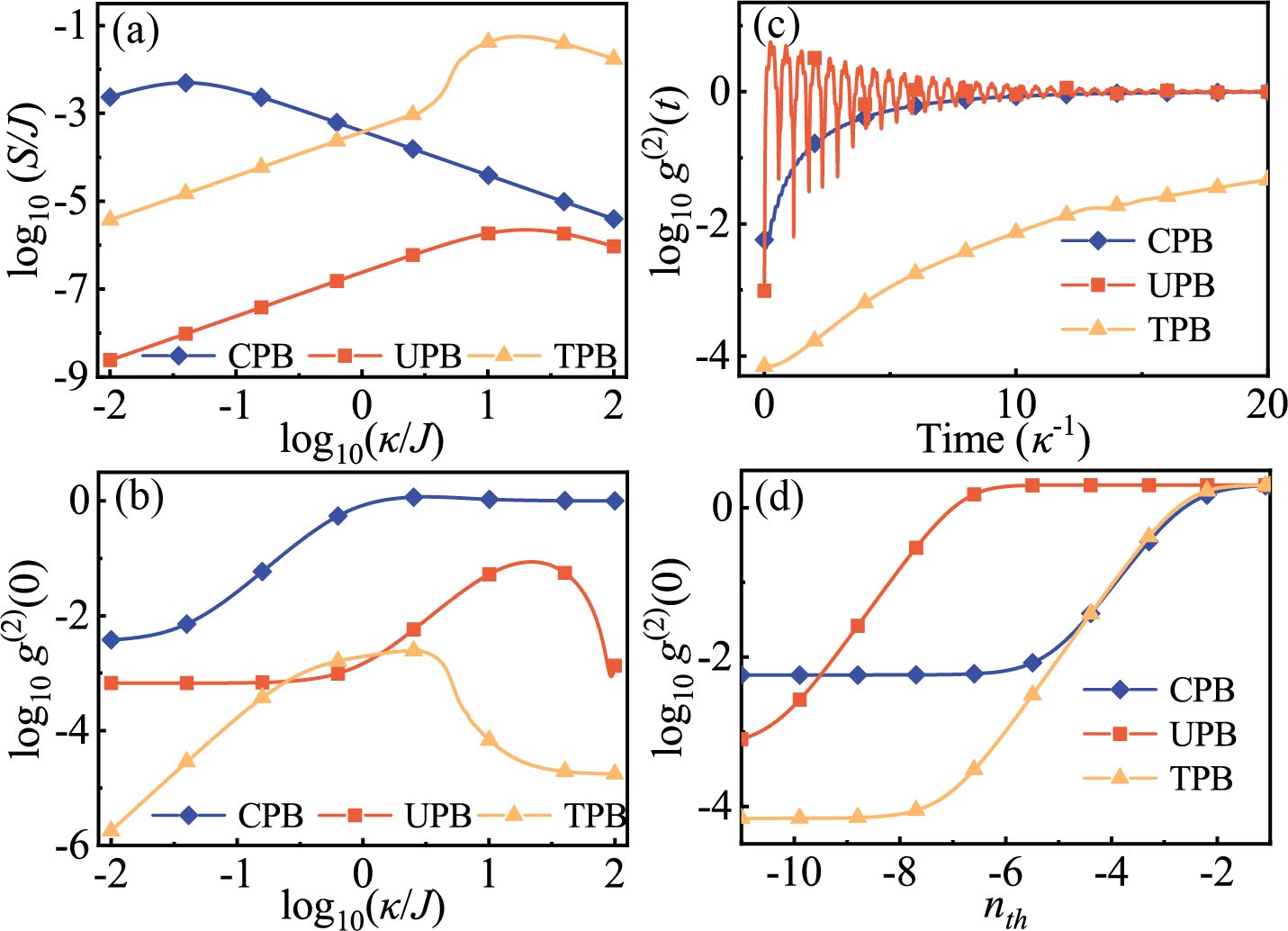}    
\caption{Performance comparison among TPB, CPB, and UPB. Scaled photon emission rate $S/J$ and (b) $\log_{10}g^{(2)}(0)$ versus $\log_{10}(\kappa/J)$. (c) Time evolution of the time-delayed second-order correlation function $\log_{10}g^{(2)}(t)$. (d) $\log_{10}g^{(2)}(0)$ as a function of the thermal occupation number $n_{\rm th}$. Optimized JC-model parameters for CPB and UPB are detailed in the Supplemental Material~\cite{SM}.}    
\label{FIG3}
\end{figure}
We next examine the influence of $\gamma$ on the PB in Fig.~\ref{FIG2}(d). The increase of the minimum $g^{(2)}(0)$ with $\gamma$ follows from the low-driving scaling relations $N \propto 1/(J^2 + \kappa\gamma)$ and $\mathcal{N}\propto \gamma^4/\Xi$, yielding $g^{(2)}(0) \propto \gamma^4(J^2 +\kappa \gamma)^2/\Xi$. This can be understood physically: a larger $\gamma$ accelerates relaxation from $|1, e_1, e_2\rangle$ to $|1, g_1, e_2\rangle$, enhancing the indirect pathway to the two-photon state and lifting $g^{(2)}(0)$. At large driving strengths, the saturated value of $N$ scales as $N \propto {\gamma}/({ C\gamma + \kappa})$ with $C=3\kappa^2/4J^2$. When $\kappa$ dominates the denominator, $N \propto \gamma$, resulting in an approximately tenfold increase as $\gamma$ varies from $10^{-3}\kappa$ to $10^{-2}\kappa$. Conversely, when $C\gamma$ dominates, the $\gamma$ factors cancel out, and the saturated $N$ levels off near $10^{-2}$, remaining nearly unchanged as $\gamma$ increases from $10^{-2}\kappa$ to $10^{-1}\kappa$. The parameter spacing between the onset of the $g^{(2)}(0)$ rise and the saturation of $N$ serves as a critical geometric indicator. Their close proximity enables the nearly simultaneous achievement of low $g^{(2)}(0)$ and high $N$. As shown in Fig.~\ref{FIG2}(d), this spacing is narrow for $\gamma=10^{-3}\kappa$ and $10^{-2}\kappa$ but becomes relatively wide for $\gamma=10^{-1}\kappa$. Thus, $\gamma=10^{-2}\kappa$ emerges as the preferable working point that successfully breaks the conventional purity-brightness trade-off, yielding $N \approx 10^{-2}$ and $g^{(2)}(0) \approx 10^{-4}$ at $\Omega=10^{-2}\kappa$.
To systematically map the full optimal operating ranges of $\Omega$ and $\gamma$, we plot $g^{(2)}(0)$ and $N$ in the $\Omega$-$\gamma$ parameter space in Figs.~\ref{FIG2}(e) and (f), respectively. The overlapping region bounded by the black and white dashed contours (indicating low $g^{(2)}(0)$ and high $N$, respectively) provides a favorable operating window with $\log_{10}(\Omega/\kappa) \in [-2.1, -1.7]$ and $\log_{10}(\gamma/\kappa) \in [-2.3, -1.7]$. Within this window, $g^{(2)}(0) \approx 10^{-4}$ stays low and non-rising, while $N \approx 10^{-2}$ remains high, enabling the realization of high-purity and high-brightness PB.

\textit{Comparison with CPB and UPB}---The brightness of a single-photon source, defined as the number of photons emitted per unit time, is measured by the emission rate $S = \kappa N$. High brightness requires both a large mean photon number $N$ and a high photon decay rate $\kappa$. Thus, for a fixed coupling strength $J$, the brightness increases with $\kappa$ provided that $N$ remains nearly constant. Benefiting from a minimally varying $N$ from the very strong ($\kappa/J=0.01$) to weak coupling ($\kappa/J=10$) regimes, the PB based on three-body interactions (TPB) achieves a maximum brightness of approximately $0.1J$ at $\log_{10}(\kappa/J) \in [1,1.5]$, as shown in Fig.~\ref{FIG3}(a). However, further increasing the dissipation causes the decrease in $N$ to outweigh the increase in $\kappa$, thereby reducing the brightness. The maximum brightness of TPB is several orders of magnitude higher than that of UPB, which suffers from a very small $N$ under comparable dissipation. It also exceeds that of CPB under strong coupling by more than an order of magnitude, because CPB's slightly higher $N$ is heavily outweighed by its two-orders-of-magnitude smaller $\kappa$. Moreover, as shown in Fig.~\ref{FIG3}(b), TPB achieves $g^{(2)}(0)$ in the range of $10^{-4}$ to $10^{-5}$ at the maximum brightness. This is far lower than that of CPB and UPB, simultaneously delivering high brightness and high purity and breaking the purity-brightness trade-off.

We then calculate the time-delayed second-order correlation function $g^{(2)}(t)=\langle\hat{a}^{\dagger}\hat{a}^{\dagger}(t)\hat{a}(t)\hat{a}\rangle/\langle\hat{a}^{\dagger}\hat{a}\rangle^{2}$. As shown in Fig.~\ref{FIG3}(c), TPB exhibits a slowly varying $g^{(2)}(t)$, whereas UPB displays rapid oscillations exceeding $1$, indicating that UPB requires a high detector time resolution to resolve the antibunching dip, while TPB offers a broad measurement window.  Finally, we investigate the effect of thermal occupation $n_{\rm th}$ on $g^{(2)}(0)$ in Fig.~\ref{FIG3}(d). Both TPB and CPB maintain strong PB ($g^{(2)}(0) \approx 10^{-2}$) up to $n_{\rm th}\approx10^{-5}$, whereas UPB degrades to this threshold already at $n_{\rm th}\approx10^{-9}$, demonstrating the superior thermal robustness of TPB. Collectively, these results demonstrate that TPB exhibits extraordinary robustness against parameter variations (including driving strength), whereas UPB is highly sensitive to experimental imperfections and imposes strict operating constraints.

\textit{Physical realization}---The blockade mechanism proposed here relies on a specific three-body interaction involving a single bosonic mode and two qubits. As a universal theoretical framework, it is applicable not only to optical photons but also to other bosonic modes such as magnons~\cite{Liu2019,Xie2020,Jin2023,Hou2024,8jy6-fp5x} and phonons~\cite{Liu2010, Xie2017, Yao2022, Zhao2020phonon, Cheng2022, Song2025}. While some schemes to realize such three-body interactions have been proposed~\cite{PhysRevLett.117.043601,zhao2025threebodyinteractionmagnonandreevsuperconductingqubit}, they have not yet been implemented in the microwave photonic platform. However, generating high-quality microwave single photons based on this mechanism is of paramount significance, as it provides a highly controllable on-chip source—a crucial building block for scalable superconducting quantum networks and distributed quantum computing~\cite{kurpiers2018deterministic, axline2018demand, PhysRevLett.125.260502}. Here, we propose a versatile architecture in a superconducting circuit, where two transmon qubits and a coplanar waveguide resonator are coherently coupled via a flux-tunable Josephson coupler~\cite{Yan2018,Neumeier2013,Jones2018,Ido2023, Hays2025PRXQuantum}. This design successfully engineers the desired three-body interaction while completely eliminating unwanted two-body couplings. The detailed superconducting circuit and Hamiltonian derivation are relegated to the Supplemental Material~\cite{SM}. By utilizing realistic experimental parameters, a three-body coupling strength of $J/2\pi \approx 10$ MHz can be achieved. When the resonator is engineered into the overcoupled regime with a decay rate of $\kappa/2\pi = 100$ MHz, this architecture yields a high-performance microwave single-photon source with an extreme purity of $g^{(2)}(0) \approx 10^{-4}$ and a remarkable emission rate of $S \approx 10^6 /\mathrm{s}$, firmly demonstrating the practical potential of our mechanism.

\textit{Conclusion}---We have proposed a novel photon blockade mechanism based on three-body interactions that fundamentally cuts off the excitation path to the two-photon state. This approach is no longer limited by coupling and driving strengths, enabling the realization of robust photon blockade over a broad parameter space. The unique mechanism allows for the nearly simultaneous realization of the lowest second-order correlation function and the highest average photon number, breaking the longstanding purity-brightness trade-off. Consequently, our scheme exhibits a significantly higher brightness than previous schemes while maintaining extreme single-photon purity. Furthermore, this blockade mechanism demonstrates remarkable robustness against thermal noise and avoids unwanted oscillations in the time-delayed correlation function. Although we present a feasible implementation in a microwave photonic platform, this mechanism is universal and can be extended to other bosonic modes (such as phonons and magnons). Our work provides a mechanism to generate a high-purity, high-brightness, and robust single-particle source, which is crucial for quantum information processing.

\textit{Note added}.—Recently, we have noticed a relevant work~\cite{lu2026highbrightnessperfectphotonblockade} that achieved PB with near-ideal purity and high mean photon number in an extended two-photon Jaynes-Cummings model with two-body and three-body interactions.

\begin{acknowledgments}
P.B.L. is supported by the National Natural Science Foundation of China under Grants No. W2411002 and No. 12375018.
\end{acknowledgments}
\bibliographystyle{apsrevlong}
\bibliography{reference}
\end{document}


\title{Supplemental Material for ``Photon blockade via three-body interactions: toward high-purity and bright single-photon sources"}
	\author{Sheng Zhao}
	\affiliation{Ministry of Education Key Laboratory for Nonequilibrium Synthesis and Modulation of Condensed Matter, Shaanxi Province Key Laboratory of Quantum Information and Quantum Optoelectronic Devices, School of Physics, Xi'an Jiaotong University, Xi'an 710049, China}
	\author{Peng-Bo Li}
	\email{lipengbo@mail.xjtu.edu.cn}
	\affiliation{Ministry of Education Key Laboratory for Nonequilibrium Synthesis and Modulation of Condensed Matter, Shaanxi Province Key Laboratory of Quantum Information and Quantum Optoelectronic Devices, School of Physics, Xi'an Jiaotong University, Xi'an 710049, China}
	
	\date{\today}
	
	\begin{abstract}
		In this Supplemental Material, we provide detailed derivations and extended analyses to support the main text. Sec.~\ref{SE1} presents the exact analytical solutions for the steady-state photon statistics within the truncated two-excitation manifold, detailing the derivation of the mean photon number and two-photon population. Sec.~\ref{SE2} provides a comprehensive analysis of both conventional and unconventional photon blockade mechanisms based on the dual-driven Jaynes-Cummings model, explicitly demonstrating their inherent purity-brightness trade-offs for comparison. Sec.~\ref{SE3} details the physical realization of the target three-body interaction in a superconducting circuit architecture, demonstrating how the pure three-body coupling is rigorously generated while residual two-body interaction is effectively suppressed.	
	\end{abstract}
	\maketitle
	\tableofcontents 
	\newpage
	\section{\label{SE1}Analytical Solution of the Steady-State Photon Statistics}
	In this section, we present a detailed analytical derivation of the steady-state photon statistics for our system. Under our proposed mechanism, the excitation of higher-order photon states is inherently and strongly suppressed, allowing us to rigorously truncate the infinite-dimensional Hilbert space to the two-excitation manifold. By reducing the master equation to a closed set of linear algebraic equations, we exactly solve the steady-state density matrix and transparent analytical expressions for both the mean photon number and the two-photon population. Finally, these analytical formulas are systematically validated against exact numerical simulations.
	
	We start with the total Hamiltonian of the three-body system consisting of a single bosonic mode and two qubits. In the rotating frame, the total Hamiltonian is given by
	\begin{equation}
		\hat{H}_{\text{tot}}=\Delta\hat{\sigma}_{1}^{+}\hat{\sigma}_{1}^{-}+\Delta\hat{\sigma}_{2}^{+}\hat{\sigma}_{2}^{-}+J(\hat{a}^{\dagger}\hat{\sigma}_{1}^{+}\hat{\sigma}_{2}^{-}+\hat{a}\hat{\sigma}_{1}^{-}\hat{\sigma}_{2}^{+})+\Omega(\hat{\sigma}_{2}^{+}+\hat{\sigma}_{2}^{-}),
	\end{equation}
	where $\Delta$ is the detuning, $J$ is the coupling strength, and $\Omega$ represents the Rabi frequency between the second qubit and the driving field. The dynamics of this open quantum system, coupled to the vacuum environment, are strictly governed by the standard Lindblad master equation:
	\begin{equation}
		\frac{\partial \rho}{\partial t} = -i[\hat{H}_{\text{tot}}, \rho] + \kappa \mathcal{D}[\hat{a}]\rho + \gamma \mathcal{D}[\hat{\sigma}_{1}^{-}]\rho + \gamma \mathcal{D}[\hat{\sigma}_{2}^{-}]\rho,
	\end{equation}
	where $\mathcal{D}[\hat{o}]\rho = \hat{o}\rho\hat{o}^{\dagger} - \frac{1}{2}(\hat{o}^{\dagger}\hat{o}\rho + \rho\hat{o}^{\dagger}\hat{o})$ denotes the standard Lindblad dissipator for a given collapse operator $\hat{o}$. Here, $\kappa$ is the photon decay rate, and $\gamma$ is the decay rate of the qubits. 
	
	To facilitate the analytical derivation, the open-system dynamics can be separated into non-Hermitian evolution and quantum jumps. By absorbing the anti-commutator parts of the dissipators into the coherent Hamiltonian, we define the effective non-Hermitian Hamiltonian as:
	\begin{equation}
		\hat{H}_{\text{eff}} = \hat{H}_{\text{tot}} - \frac{i}{2} \kappa \hat{a}^{\dagger}\hat{a} - \frac{i}{2} \gamma \hat{\sigma}_{1}^{+}\hat{\sigma}_{1}^{-} - \frac{i}{2} \gamma \hat{\sigma}_{2}^{+}\hat{\sigma}_{2}^{-}.
	\end{equation}
	
	Consequently, the master equation can be equivalently rewritten in terms of $\hat{H}_{\text{eff}}$ and the remaining quantum jump terms:
	\begin{equation}
		\frac{\partial \rho}{\partial t} = -i(\hat{H}_{\text{eff}}\rho - \rho \hat{H}_{\text{eff}}^{\dagger}) + \kappa \hat{a} \rho \hat{a}^{\dagger} + \gamma \hat{\sigma}_{1}^{-} \rho \hat{\sigma}_{1}^{+} + \gamma \hat{\sigma}_{2}^{-} \rho \hat{\sigma}_{2}^{+}.
	\end{equation}
	
	To obtain analytical solutions, the infinite-dimensional Hilbert space is rigorously truncated to the two-photon subspace, since the excitation of higher Fock states is strongly suppressed under our mechanism. Within this truncated 10-dimensional subspace, we define the complete basis set $\mathcal{B} = \{ |v_0\rangle, |v_1\rangle, \dots, |v_9\rangle \}$ corresponding to the physical states of the system in the following order:
	\begin{align}
		|v_0\rangle &= |0, g_1, g_2\rangle, \quad |v_1\rangle = |0, e_1, g_2\rangle, \quad  |v_2\rangle = |0, g_1, e_2\rangle, \quad  |v_3\rangle = |0, e_1, e_2\rangle, \quad 
		|v_4\rangle = |1, g_1, g_2\rangle,\nonumber\\
		|v_5\rangle &= |1, e_1, g_2\rangle, \quad 
		|v_6\rangle = |1, g_1, e_2\rangle, \quad  |v_7\rangle = |1, e_1, e_2\rangle, \quad 
		|v_8\rangle = |2, g_1, g_2\rangle, \quad  |v_9\rangle = |2, e_1, g_2\rangle.
	\end{align}
	Under this specific representation, the steady-state density matrix $\rho_s$ is explicitly expressed as a $10 \times 10$ matrix, whose elements are defined by $\rho_{ij} = \langle v_i | \rho_s | v_j \rangle$ ($i, j = 0, 1, \dots, 9$).
	For notational simplicity and compactness in the matrix representation, we introduce the complex detuning by absorbing the qubit decay rate into a newly defined parameter $\tilde{\Delta} = \Delta - \frac{i}{2}\gamma$.
	Consequently, the effective non-Hermitian Hamiltonian $\hat{H}_{\text{eff}}$ can be explicitly recast into the following matrix form:
	\setcounter{MaxMatrixCols}{10}
	\begin{align}
		\hat{H}_{\text{eff}} = \begin{pmatrix}
			0 & 0 & \Omega  & 0 & 0 & 0 & 0 & 0 & 0 & 0 \\
			0 & \tilde{\Delta} & 0  & \Omega & 0 & 0 & 0 & 0 & 0 & 0  \\
			\Omega & 0 & \tilde{\Delta}  & 0 & 0 & J & 0  & 0 & 0 & 0 \\
			0 & \Omega & 0 & 2\tilde{\Delta} & 0 & 0  & 0  & 0 & 0 & 0\\
			0 & 0 & 0 & 0 & -\frac{i}{2}\kappa & 0  & \Omega  & 0 & 0 & 0\\
			0 & 0 & J & 0 & 0 &  \tilde{\Delta}-\frac{i}{2}\kappa & 0  & \Omega & 0 & 0 \\
			0 & 0 & 0 & 0 &  \Omega & 0 & \tilde{\Delta}-\frac{i}{2}\kappa &  0 & 0 & \sqrt{2}J \\
			0 & 0 & 0 &  0 & 0 & \Omega &  0 &  2\tilde{\Delta}-\frac{i}{2}\kappa & 0 & 0  \\
			0 & 0 & 0 & 0 & 0 &  0 & 0 & 0 &  -{i}\kappa  & 0 \\
			0 & 0 & 0 & 0 & 0 &  0 & \sqrt{2}J & 0 &  0 & \tilde{\Delta}-{i}\kappa  
		\end{pmatrix}.
	\end{align}
	
	To complete the transformation of the Lindblad master equation into a fully solvable algebraic system, the quantum jump terms must also be projected onto this subspace. Specifically, the cavity jump term $\kappa \hat{a} \rho_s \hat{a}^{\dagger}$ is strictly mapped to:
	\begin{align}
		\mathcal{J}_{\kappa}[\rho_s] = \kappa
		\begin{pmatrix}
			\rho_{44} & \rho_{45} & \rho_{46} & \rho_{47} & \sqrt{2}\rho_{48} & \sqrt{2}\rho_{49} & 0 & 0 & 0 & 0 \\
			\rho_{54} & \rho_{55} & \rho_{56} & \rho_{57} & \sqrt{2}\rho_{58} & \sqrt{2}\rho_{59} & 0 & 0 & 0 & 0 \\
			\rho_{64} & \rho_{65} & \rho_{66} & \rho_{67} & \sqrt{2}\rho_{68} & \sqrt{2}\rho_{69} & 0 & 0 & 0 & 0 \\
			\rho_{74} & \rho_{75} & \rho_{76} & \rho_{77} & \sqrt{2}\rho_{78} & \sqrt{2}\rho_{79} & 0 & 0 & 0 & 0 \\
			\sqrt{2}\rho_{84} & \sqrt{2}\rho_{85} & \sqrt{2}\rho_{86} & \sqrt{2}\rho_{87} & 2\rho_{88} & 2\rho_{89} & 0 & 0 & 0 & 0 \\
			\sqrt{2}\rho_{94} & \sqrt{2}\rho_{95} & \sqrt{2}\rho_{96} & \sqrt{2}\rho_{97} & 2\rho_{98} & 2\rho_{99} & 0 & 0 & 0 & 0 \\
			0 & 0 & 0 & 0 & 0 & 0 & 0 & 0 & 0 & 0 \\
			0 & 0 & 0 & 0 & 0 & 0 & 0 & 0 & 0 & 0 \\
			0 & 0 & 0 & 0 & 0 & 0 & 0 & 0 & 0 & 0 \\
			0 & 0 & 0 & 0 & 0 & 0 & 0 & 0 & 0 & 0 
		\end{pmatrix}.
	\end{align}
	Similarly, the qubit jump terms $\gamma \hat{\sigma}_{1}^{-} \rho_s \hat{\sigma}_{1}^{+} + \gamma \hat{\sigma}_{2}^{-} \rho_s \hat{\sigma}_2^+$ are jointly expressed as:
	\begin{align}
		\mathcal{J}_{\gamma}[\rho_s] = \gamma
		\begin{pmatrix}
			\rho_{11}+\rho_{22} & \rho_{23} & \rho_{13} & 0 & \rho_{15}+\rho_{26} & \rho_{27} & \rho_{17} & 0 & \rho_{19} & 0 \\
			\rho_{32} & \rho_{33} & 0 & 0 & \rho_{36} & \rho_{37} & 0 & 0 & 0 & 0 \\
			\rho_{31} & 0 & \rho_{33} & 0 & \rho_{35} & 0 & \rho_{37} & 0 & \rho_{39} & 0 \\
			0 & 0 & 0 & 0 & 0 & 0 & 0 & 0 & 0 & 0 \\
			\rho_{51}+\rho_{62} & \rho_{63} & \rho_{53} & 0 & \rho_{55}+\rho_{66} & \rho_{67} & \rho_{57} & 0 & \rho_{59} & 0 \\
			\rho_{72} & \rho_{73} & 0 & 0 & \rho_{76} & \rho_{77} & 0 & 0 & 0 & 0 \\
			\rho_{71} & 0 & \rho_{73} & 0 & \rho_{75} & 0 & \rho_{77} & 0 & \rho_{79} & 0 \\
			0 & 0 & 0 & 0 & 0 & 0 & 0 & 0 & 0 & 0 \\
			\rho_{91} & 0 & \rho_{93} & 0 & \rho_{95} & 0 & \rho_{97} & 0 & \rho_{99} & 0 \\
			0 & 0 & 0 & 0 & 0 & 0 & 0 & 0 & 0 & 0 
		\end{pmatrix}.
	\end{align}
	
	By substituting these explicit matrix representations into the steady-state condition $\partial\rho_s/\partial t = 0$, the Lindblad master equation is rigorously expanded into the following matrix equation:
	\begin{align}
		-i (\hat{H}_{\text{eff}} \rho_s - \rho_s \hat{H}_{\text{eff}}^\dagger) + \mathcal{J}_{\kappa}[\rho_s] + \mathcal{J}_{\gamma}[\rho_s] = \mathbf{0},
	\end{align}
	where $\mathbf{0}$ denotes a $10 \times 10$ zero matrix. By solving this matrix equation, the steady-state density matrix $\rho_s$ of the system can be exactly obtained. With the exact $\rho_s$ in hand, the photon statistics of the system can be characterized by the equal-time second-order correlation function, which is defined as
	\begin{align}
		g^{(2)}(0) = \frac{\langle \hat{a}^\dagger \hat{a}^\dagger \hat{a} \hat{a} \rangle}{\langle \hat{a}^\dagger \hat{a} \rangle^2}=\frac{\text{Tr}(\hat{a}^\dagger \hat{a}^\dagger \hat{a} \hat{a} \rho_s) }{[\text{Tr}(\hat{a}^\dagger \hat{a} \rho_s)]^2}.
	\end{align}
	To explicitly analyze this correlation function, our primary task is to extract the steady-state mean photon number (for the denominator) and the two-photon population (for the numerator). First, the steady-state mean photon number $N = \langle \hat{a}^\dagger \hat{a} \rangle$ can be directly evaluated as
	\begin{align}
		N = \langle \hat{a}^\dagger \hat{a} \rangle = \text{Tr}(\hat{a}^\dagger \hat{a} \rho_s) \approx \rho_{55}.
	\end{align}
	This approximation is physically rigorous because the population of the single-photon manifold is overwhelmingly dominated by the state $|1, e_1, g_2\rangle$. Specifically, the states $|1, g_1, g_2\rangle$ and $|1, g_1, e_2\rangle$ remain largely unpopulated, as they can only be weakly accessed through the qubit dissipation channels. Furthermore, reaching the state $|1, e_1, e_2\rangle$ requires an additional excitation from $|1, e_1, g_2\rangle$, rendering its population small. Consequently, the condition $\rho_{55} \gg \rho_{44}, \rho_{66}, \rho_{77}$ is satisfied, allowing us to safely neglect the minor contributions from other single-photon states. Moreover, since the populations of the two-photon states are orders of magnitude smaller, the mean photon number is governed by $\rho_{55}$. To obtain a concise and physically transparent analytical expression, we assume that the driving strength $\Omega$ and the qubit decay rate $\gamma$ are relatively small. By systematically neglecting the higher-order terms of $\Omega$ and $\gamma$, the steady-state mean photon number can be simplified as
	\begin{align}
		N \approx \frac{J^2\Omega^2}{(J^2-\Delta^2)^2+\Delta^2\kappa^2/4+J^2\kappa\gamma+\alpha\Omega^2},
	\end{align}
	where $\alpha = J^2\kappa/\gamma+3\kappa^2/4$. Similarly, the second-order moment can be directly evaluated as:
	\begin{align}
		\langle \hat{a}^\dagger \hat{a}^\dagger \hat{a} \hat{a} \rangle = \text{Tr}(\hat{a}^\dagger \hat{a}^\dagger \hat{a} \hat{a} \rho_s) = 2(\rho_{88} + \rho_{99}) = 2\mathcal{N},
	\end{align}
	where $\mathcal{N}$ denotes the steady-state two-photon population. Furthermore, the two-photon manifold is predominantly occupied by the state $|2, e_1, g_2\rangle$. Since reaching the state $|2, g_1, g_2\rangle$ requires an additional qubit dissipation process from $|2, e_1, g_2\rangle$, its plpulation is negligibly small. Therefore, the condition $\rho_{99} \gg \rho_{88}$ is well justified, and we obtain the approximated two-photon population as $\mathcal{N} \approx \rho_{99}$. 
	
	To extract physically transparent formulas for $\mathcal{N}$, we again apply the weak-driving and small-decay approximations. However, because the exact algebraic expression for $\mathcal{N}$ remains exceedingly cumbersome, we further evaluate it in two distinct asymptotic limits depending on the magnitude of the coupling strength $J$: the weak-coupling limit ($J \ll \kappa$) and the strong-coupling limit ($J \gg \kappa$). By systematically retaining only the leading-order contributions in each respective limit, the simplified asymptotic expressions are obtained as follows. In the weak-coupling limit ($J \ll \kappa$), $\mathcal{N}$ reduces to:
	\begin{equation}
		\mathcal{N} \approx  \frac{21J^4 \gamma^2 [ (J^2 \gamma^2+\mathcal{M}_{1\Delta})\Omega^4  +  (1.5J^2 + 6.5\gamma^2)\Omega^6 ]}{\kappa^4 ( \Xi+ 28 J^2 \kappa^2 \gamma^4 \Omega^2 + \Theta \Omega^4 )+\mathcal{M}_{2\Delta}},
	\end{equation}
	where 
	\begin{align}
		\mathcal{M}_{1\Delta} &= \Delta^8\gamma\kappa^{-5}+\Delta^{10}(0.07\gamma\kappa^{-7}+0.5\gamma^2\kappa^{-8}),\nonumber\\
		\mathcal{M}_{2\Delta} &= \Delta^{14} (20 \gamma^2 \kappa^{-2} + 230 \gamma^3 \kappa^{-3}) + \Delta^{16} (17 J^2 \gamma \kappa^{-5} + 130 J^2 \gamma^2 \kappa^{-6} + 4 \gamma^2 \kappa^{-4} + 40 \gamma^3 \kappa^{-5} + 170 \gamma^4 \kappa^{-6}),
	\end{align}
	with parameters $\Xi=2J^{6}\kappa\gamma^{3}+42J^{6}\gamma^{4}+J^{4}\kappa^{2}\gamma^{4}$ and $\Theta=2J^{4}\kappa^{2}+1.5J^{2}\kappa^{3}\gamma+500J^{2}\kappa\gamma^{3}+6\kappa^{3}\gamma^{3}+80\kappa^{2}\gamma^{4}$. 
	
	Conversely, in the strong-coupling limit ($J \gg \kappa$), it is approximated as
	\begin{align}
		\mathcal{N} \approx \frac{2J^6 \Omega^4 \gamma}{5\kappa^3} \frac{\big[ \Lambda_{1\Delta} + \Lambda_{2\Delta} + 7 J^6 \kappa^2 \gamma^2 \big]}{\big[ \Lambda_{3\Delta} + \Lambda_{4\Delta} + 25 J^{14} \kappa \gamma^3 \big]},
	\end{align}
	where
	\begin{align}
		\Lambda_{1\Delta} =& \Delta^4 J^4 (22 \kappa^2 + 158 \kappa \gamma + 431 \gamma^2) + \Delta^2 J^6 \kappa (3 \kappa + 21 \gamma),\nonumber \\
		\Lambda_{2\Delta} =& \Delta^8 (26 \kappa^2 + 189 \kappa \gamma + 506 \gamma^2) - \Delta^6 J^2 (52 \kappa^2 + 372 \kappa \gamma + 567 \gamma^2),\nonumber \\
		\Lambda_{3\Delta} =&   \Delta^8 J^8 (153 J^2 + 91 \kappa^2 + 806 \kappa \gamma) 
		- \Delta^6 J^{10} (75 J^2 + 208 \kappa^2) + \Delta^4 J^{12} (15 J^2 + 38 \kappa^2 + 370 \kappa \gamma), \nonumber\\
		\Lambda_{4\Delta} =& \Delta^{16} (4 J^2 + \kappa^2 + 10 \kappa \gamma) + \Delta^{14} J^2 (-30 J^2 + 11 \kappa^2 + 92 \kappa \gamma) \nonumber \\
		&+ \Delta^{12} J^4 (97 J^2 - 260 \Omega^2 - 32 \kappa^2 - 302 \kappa \gamma)+\Delta^{10} J^6 (-165 J^2 + 391 \Omega^2).
	\end{align}
	
	\begin{figure}[tb]
		\centering
		\includegraphics[width=\linewidth]{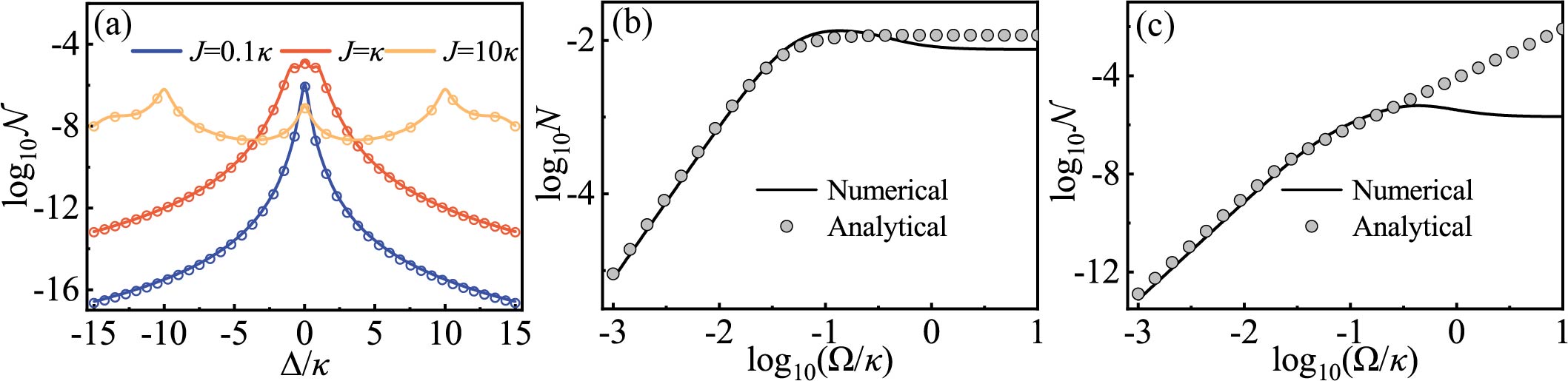} 
		\caption{Comparison between the analytical results (symbols) and numerical simulations (solid lines). (a) The steady-state two-photon population $\mathcal{N}$ as a function of the detuning $\Delta$ for different coupling strengths $J$, with $\Omega = 0.1\kappa$ and $\gamma = 0.1\kappa$. (b) The mean photon number $N$ and (c) the two-photon population $\mathcal{N}$ as functions of the driving strength $\Omega$ at the optimal detuning condition $\Delta = 0$, with $J=0.1\kappa$ and $\gamma = 0.1\kappa$.}
		\label{F1}
	\end{figure}
	
	To further validate the accuracy of our theoretical framework, we compare the analytical and numerical results for the two-photon population $\mathcal{N}$. As shown in Fig.~\ref{F1}(a), the analytical results are in excellent agreement with the numerical results across the considered parameter space. It is noteworthy that under the optimal detuning condition, the two-photon population $\mathcal{N}$ also attains a local maximum. However, because the mean photon number $N$ is simultaneously maximized and the drastic enhancement of its squared term $N^2$ exerts a much more dominant impact, the steady-state second-order correlation function $g^{(2)}(0) \propto 2\mathcal{N}/N^2$ ultimately achieves its minimum value. In this respect, our scheme is phenomenologically similar to the conventional photon blockade, as both achieve a minimized $g^{(2)}(0)$ by maximizing the mean photon number $N$ via single-photon resonance. However, the underlying physical mechanisms for suppressing two-photon excitations are fundamentally distinct: while CPB relies on strong energy anharmonicity to energetically block two-photon transitions, our approach achieves this by strictly cutting off the coherent excitation path.
	
	Furthermore, in Figs.~\ref{F1}(b) and \ref{F1}(c), we systematically investigate the validity range of our analytical solutions as the driving strength $\Omega$ increases. As explicitly illustrated, as long as the driving strength is not too strong (e.g., $\log_{10}(\Omega/\kappa) \le -0.5$), the analytical solutions for both $N$ and $\mathcal{N}$ perfectly match the exact numerical results. This strictly justifies our theoretical analysis in the main text, demonstrating that our analytical formulas can accurately capture the scaling behaviors of the mean photon number and the steady-state second-order correlation function with respect to the driving strength. However, when $\Omega$ becomes relatively large, the higher-order terms of $\Omega$ that were neglected in our derivation begin to have a substantial impact, leading to noticeable deviations. Specifically, while the mean photon number $N$ exhibits only a minor deviation, the two-photon population $\mathcal{N}$ shows a much more significant discrepancy in this strong-driving regime.
	\section{\label{SE2} Conventional and Unconventional Photon Blockade}
	As discussed in the main text, we have systematically compared our proposed mechanism with the conventional photon blockade (CPB) and unconventional photon blockade (UPB). To provide a clear and self-contained physical picture, in this section we present the standard implementations and detailed analytical solutions for both the CPB and UPB based on the Jaynes-Cummings (JC) model. Furthermore, the trade-off between the second-order correlation function $g^{(2)}(0)$ and the mean photon number $N_c$ for these standard blockade mechanisms is also clearly demonstrated.
	
	\subsection{Analytical Solution of the Dual-Driven JC Model}
	Specifically, we consider a dual-driven JC model consisting of a single-mode cavity and a two-level qubit. Let $\omega_c$ and $\omega_q$ denote the resonance frequencies of the cavity mode and the qubit transition, respectively. Here, we focus on the strict resonant case where $\omega_c = \omega_q \equiv \omega_0$. Both the cavity mode and the qubit are coherently driven by external driving fields with the same driving frequency $\omega_l$. The driving strengths applied to the cavity and the qubit are quantified by $\Omega_c$ and $\Omega_q$, respectively. Notably, the primary purpose of introducing this dual-driving configuration is to construct the additional excitation pathways requisite for the destructive quantum interference in UPB. In contrast, for CPB, which fundamentally relies on the anharmonicity of the energy spectrum, the specific choice between single and dual driving has negligible effect on the blockade.
	
	In the laboratory frame, the total Hamiltonian of the driven system exhibits explicit time dependence due to the external driving fields. To eliminate this time dependence and simplify the analytical description, we apply a unitary transformation to a rotating frame with respect to the driving frequency $\omega_l$, defined by the unitary operator $\hat{U}(t) = \exp[-i\omega_l t (\hat{a}^\dagger\hat{a} + \hat{\sigma}_+\hat{\sigma}_-) ]$. Then, in this rotating frame, the total Hamiltonian of the system can be expressed as:
	\begin{align}
		\hat{H}_{\text{JC}} = \Delta_0 ( \hat{a}^\dagger \hat{a} + \hat{\sigma}_+ \hat{\sigma}_- ) + G(\hat{a}\hat{\sigma}_+ + \hat{a}^\dagger \hat{\sigma}_-) 
		+ \Omega_c(\hat{a}^\dagger + \hat{a}) + \Omega_q(\hat{\sigma}_+ + \hat{\sigma}_-).
	\end{align}
	Here, $\hat{a}^\dagger$ ($\hat{a}$) is the creation (annihilation) operator for the cavity mode, and $\hat{\sigma}_+$ ($\hat{\sigma}_-$) is the raising (lowering) operator for the two-level qubit. Moreover, $\Delta_0 = \omega_0 - \omega_l$ denotes the frequency detuning of both the cavity and the qubit from the driving fields, and $G$ is the coupling strength of the JC interaction. 
	
	Taking into account the inevitable coupling to the environment, the full dissipative dynamics of this open quantum system are governed by the Lindblad master equation:
	\begin{align}
		\frac{\partial\rho}{\partial t} = -i[\hat{H}_{\text{JC}}, \rho] + \kappa_a \mathcal{D}[\hat{a}]\rho + \gamma_q \mathcal{D}[\hat{\sigma}_-]\rho,
	\end{align}
	where $\mathcal{D}[\hat{o}]\rho = \hat{o}\rho\hat{o}^\dagger - \frac{1}{2}\{\hat{o}^\dagger\hat{o}, \rho\}$ is the standard Lindblad dissipator for a collapse operator $\hat{o}$ with decay rate $\kappa_a$ (for the cavity) or $\gamma_q$ (for the qubit).
	
	In the weak-driving limit, the excitation of higher-order photon states is suppressed, allowing us to safely truncate the Hilbert space to the two-excitation manifold. Furthermore, following standard theoretical treatments~\cite{RevModPhys.70.101, PhysRevA.100.062131}, the quantum jump terms ($\kappa_a\hat{a}\rho\hat{a}^\dagger$ and $\gamma_q\hat{\sigma}_-\rho\hat{\sigma}_+$) can be safely neglected. This simplification is highly valid here because these jump terms merely represent incoherent feeding from higher-excitation states with negligibly small populations, making their feedback contributions to the low-excitation dynamics insignificant. 
	
	Crucially, we emphasize a fundamental distinction between these standard blockade mechanisms and our proposed scheme. In the standard CPB and UPB frameworks, the two-photon populations are primarily governed by coherent driving pathways, allowing the blockade physics to be perfectly captured by an wavevector $|\psi\rangle$ evolving under a non-Hermitian Hamiltonian. In stark contrast, our proposed mechanism strictly relies on the weak quantum jump channels (i.e., qubit dissipation) to populate the two-photon state. Consequently, the effective non-Hermitian method is fundamentally inapplicable to our scheme, necessitating the exact steady-state solutions derived in the Sec.~\ref{SE1}.
	
	For the traditional JC model considered in this section, however, we can safely proceed with the effective non-Hermitian Schrödinger equation:
	\begin{align}
		i \frac{\partial |\psi\rangle}{\partial t} = \hat{H}_{\text{non}} |\psi\rangle,
	\end{align}
	where the effective non-Hermitian Hamiltonian is given by:
	\begin{align}
		\hat{H}_{\text{non}} = \tilde{\Delta}_c \hat{a}^\dagger \hat{a} + \tilde{\Delta}_q \hat{\sigma}_+ \hat{\sigma}_- + G(\hat{a}\hat{\sigma}_+ + \hat{a}^\dagger \hat{\sigma}_-) 
		+ \Omega_c(\hat{a}^\dagger + \hat{a}) + \Omega_q(\hat{\sigma}_+ + \hat{\sigma}_-).
	\end{align}
	For notational compactness, we have absorbed the decay rates into the definitions of the complex detunings $\tilde{\Delta}_c = \Delta_0 - i\kappa_a/2$ and $\tilde{\Delta}_q = \Delta_0 - i\gamma_q/2$.
	
	The state vector $|\psi\rangle$ within the truncated two-excitation manifold is expanded as:
	\begin{align}
		|\psi\rangle = C_{0,g}|0, g\rangle + C_{0,e}|0, e\rangle + C_{1,g}|1, g\rangle + C_{1,e}|1, e\rangle + C_{2,g}|2, g\rangle,
	\end{align}
	where $|n, g(e)\rangle$ denotes the state with $n$ photons and the qubit in the ground (excited) state.
	
	Substituting this truncated state vector into the non-Hermitian Schrödinger equation and applying the weak-driving approximation $C_{0,g} \approx 1 \gg C_{0,e}, C_{1,g}$, we obtain the following set of coupled differential equations:
	\begin{align}
		i\dot{C}_{0,e} &= \Omega_q + \tilde{\Delta}_q C_{0,e} + G C_{1,g}, \nonumber \\
		i\dot{C}_{1,g} &= \Omega_c + G C_{0,e} + \tilde{\Delta}_c C_{1,g}, \nonumber \\
		i\dot{C}_{1,e} &= \Omega_c C_{0,e} + \Omega_q C_{1,g} + (\tilde{\Delta}_c + \tilde{\Delta}_q) C_{1,e} + \sqrt{2}G C_{2,g}, \nonumber \\
		i\dot{C}_{2,g} &= \sqrt{2}\Omega_c C_{1,g} + \sqrt{2}G C_{1,e} + 2\tilde{\Delta}_c C_{2,g}.
	\end{align}
	
	By enforcing the steady-state condition $\dot{C}_{ij} = 0$, the closed-form solutions for all the relevant probability amplitudes are analytically derived as:
	\begin{align}
		C_{0,e} &= \frac{\Omega_q \tilde{\Delta}_c - \Omega_c G}{C}, \nonumber \\
		C_{1,g} &= \frac{\Omega_c \tilde{\Delta}_q - \Omega_q G}{C}, \nonumber \\
		C_{1,e} &= \frac{\Omega_c^2}{CD} \left[ G(\tilde{\Delta}_c + \tilde{\Delta}_q) - \lambda \left(\tilde{\Delta}_c(\tilde{\Delta}_c + \tilde{\Delta}_q) + G^2\right) + \lambda^2 G \tilde{\Delta}_c \right], \nonumber \\
		C_{2,g} &= \frac{\sqrt{2}\Omega_c^2}{2CD} \left[ -\tilde{\Delta}_q (\tilde{\Delta}_c + \tilde{\Delta}_q) + 2\lambda G (\tilde{\Delta}_c + \tilde{\Delta}_q) - (1+\lambda^2)G^2 \right],
	\end{align}
	where we have defined the auxiliary parameters:
	\begin{align}
		C &= G^2 - \tilde{\Delta}_c \tilde{\Delta}_q, \nonumber \\
		D &= \tilde{\Delta}_c(\tilde{\Delta}_c + \tilde{\Delta}_q) - G^2, \nonumber \\
		\lambda &= \Omega_q / \Omega_c.
	\end{align}
	Then the equal-time second-order correlation function $g^{(2)}(0)$ can be written as
	\begin{align}
		g^{(2)}(0) &=\frac{\langle \hat{a}^{\dagger}\hat{a}^{\dagger}\hat{a}\hat{a} \rangle}{\langle \hat{a}^{\dagger}\hat{a} \rangle^2}= \frac{2|C_{2,g}|^2}{(|C_{1,g}|^2 + |C_{1,e}|^2 + 2|C_{2,g}|^2)^2} \nonumber \\
		&\approx \frac{2|C_{2,g}|^2}{|C_{1,g}|^4} = \frac{|B|^2 |C|^2}{A^2 |D|^2},
	\end{align}
	where the relevant terms are explicitly given by:
	\begin{align}
		|C_{1,g}|^2 &= \frac{\Omega_c^2 A}{|C|^2}, \quad |C_{2,g}|^2 = \frac{\Omega_c^4 |B|^2}{2|C|^2|D|^2}, \nonumber \\
		|A|^2 &= \left[ (\Delta_0 - \lambda G)^2 + \frac{\gamma_q^2}{4} \right]^2, \nonumber \\
		|B|^2 &= \left[ -2\Delta_0^2 + 4\lambda G \Delta_0 - (1+\lambda^2)G^2 + \frac{\gamma_q(\kappa_a+\gamma_q)}{4} \right]^2 + \left[ \Delta_0\frac{\kappa_a+3\gamma_q}{2} - \lambda G(\kappa_a+\gamma_q) \right]^2, \nonumber \\
		|C|^2 &= \left[ G^2 - \Delta_0^2 + \frac{\kappa_a\gamma_q}{4} \right]^2 + \Delta_0^2 \frac{(\kappa_a+\gamma_q)^2}{4}, \nonumber \\
		|D|^2 &= \left[ 2\Delta_0^2 - G^2 - \frac{\kappa_a(\kappa_a+\gamma_q)}{4} \right]^2 + \Delta_0^2 \frac{(3\kappa_a+\gamma_q)^2}{4}.
		\label{eqb}
	\end{align}
	
	\subsection{Conventional Photon Blockade}
	
	We first investigate the realization of the conventional photon blockade (CPB). Physically, the CPB relies on the strong anharmonicity of the energy spectrum induced by the strong coherent coupling ($G \gg \kappa_a, \gamma_q$) between the cavity mode and the qubit. This interaction splits the single-excitation manifold into two dressed states, $|\pm\rangle = (|1,g\rangle \pm |0,e\rangle)/\sqrt{2}$, with an energy splitting of $2G$. When the driving field is exactly resonant with one of these dressed states, the single-photon excitation is resonantly enhanced. In contrast, the two-excitation manifold exhibits a different splitting of $2\sqrt{2}G$. This inherent anharmonicity introduces a significant energy mismatch the two-photon transition, thereby strongly suppressing the subsequent absorption of a second photon.
	
	Mathematically, this resonance dictates that the single-photon population $\langle \hat{a}^{\dagger}\hat{a} \rangle \approx |C_{1,g}|^2$ must be maximized. Based on our exact analytical solutions, the maximum of $|C_{1,g}|^2 \propto 1/|C|^2$ is naturally achieved when the denominator $|C|^2$ approaches its minimum $(|C|^2 \rightarrow 0)$. Given that the CPB strictly operates in the strong-coupling regime ($G \gg \kappa_a, \gamma_q$), the terms containing the dissipation rates $\kappa_a$ and $\gamma_q$ have a negligibly small effect. Consequently, by setting the dominant part of the denominator ($G^2 - \Delta_0^2$) to zero and safely ignoring the minuscule dissipation-induced terms, we straightforwardly obtain the resonance condition:
	\begin{align}
		\Delta_0 = \pm G. \label{eq:CPB_condition}
	\end{align}
	
	Under this optimal detuning condition, the strong suppression of the two-photon probability relative to the resonantly enhanced single-photon probability naturally ensures that the second-order correlation function $g^{(2)}(0) = {|B|^2 |C|^2}/(A^2 |D|^2)$ attains its minimum value. Thus, $\Delta_0 = \pm G$ leads to $C \rightarrow 0$, which provides the exact solution for the CPB.
	
\begin{figure}[tb]
		\centering
		\includegraphics[width=0.9\linewidth]{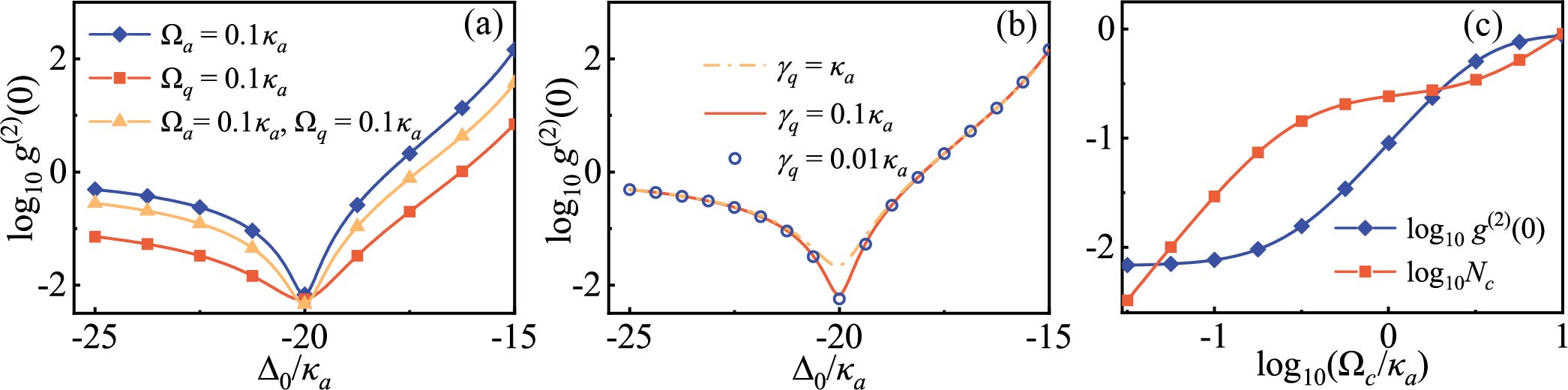}
		\caption{Performance of the CPB in the strong-coupling regime. (a) Steady-state second-order correlation function $g^{(2)}(0)$ as a function of the normalized detuning $\Delta_0/\kappa_a$ for three different driving configurations with $\gamma_q = 0.1\kappa_a$ and $G = 20 \kappa_a$. (b) $g^{(2)}(0)$ versus the normalized detuning $\Delta_0/\kappa_a$ for varying qubit dissipation rates with $\Omega_c = 0.1\kappa_a$ and $G = 20 \kappa_a$. (c) $g^{(2)}(0)$ and the mean photon number $N_c$ as functions of the normalized driving strength $\Omega_c/\kappa_a$ at the optimal CPB resonance with $\gamma_q = 0.1\kappa_a$ and $G = 20 \kappa_a$.}
		\label{fig2}
	\end{figure}

	In Fig.~\ref{fig2}(a), we plot the steady-state second-order correlation function $g^{(2)}(0)$ as a function of the detuning $\Delta_0$ for three different driving configurations: single-cavity, single-qubit, and dual driving. Off-resonance, the curves exhibit distinct slopes and varying behaviors due to the different multiphoton excitation pathways and resulting quantum interference introduced by each specific driving term. However, exactly at the optimal blockade position ($|\Delta_0| = G$), all three curves perfectly converge to the same minimum value. This clearly indicates that the specific choice of driving fields has a negligible impact on the ultimate quality of the CPB, which is fundamentally governed by the anharmonicity of the strongly coupled system.
	
	In Fig.~\ref{fig2}(b), we examine the influence of the qubit dissipation rate $\gamma_q$ on the CPB by plotting $g^{(2)}(0)$ as a function of the detuning $\Delta_0$ for different values of $\gamma_q$. According to our analytical expressions, at the resonance condition, the correlation function is directly proportional to $|C|^2 \propto (\kappa_a+\gamma_q)^2$. As clearly shown in the figure, when $\gamma_q$ is reduced from $\kappa_a$ to $0.1\kappa_a$, the minimum value of $g^{(2)}(0)$ decreases. However, as $\gamma_q$ is further reduced to $0.01\kappa_a$, the correlation function does not decrease indefinitely but rather saturates to a finite lower bound. This saturation visually confirms that the CPB is ultimately limited by the cavity dissipation $\kappa_a$. This behavior is in stark contrast to our newly proposed mechanism, where $g^{(2)}(0)$ can continuously approach zero as the qubit dissipation is minimized, free from the cavity-dissipation limitation.
	
	Furthermore, Fig.~\ref{fig2}(c) depicts the variations of $g^{(2)}(0)$ and the mean photon number $N_c$ as the driving strength $\Omega_c$ increases at the optimal CPB resonance. While a stronger driving field significantly enhances the mean photon number $N_c$, it simultaneously degrades the blockade quality ($g^{(2)}(0)$ increases), visually demonstrating the inherent trade-off between single-photon purity and source brightness in the standard CPB.
	
	Finally, we briefly outline the parameter settings used for the CPB results depicted in Fig.~3 of the main text. As derived analytically, the optimal detuning is set to the dressed-state resonance, $|\Delta_0| = G$. Furthermore, the CPB tolerance to the driving strength scales positively with the coupling strength. Therefore, to ensure a sufficiently small second-order correlation function and maintain a high-quality blockade, the driving strength is conservatively chosen as $\Omega_c = 0.01G$. For the remaining parameters, such as the qubit dissipation rate, we strictly maintain $\gamma_q = 0.01\kappa_a$. This ensures complete consistency with the settings of our newly proposed mechanism, thereby allowing for a direct and fair physical comparison.
	
	\subsection{Unconventional Photon Blockade}
	
	Unlike the CPB that relies on strong energy anharmonicity, the unconventional photon blockade (UPB) originates from destructive quantum interference. In a coupled photon-qubit system, UPB can be effectively realized by engineering multiple distinct excitation pathways. 
	
	By precisely tuning the system parameters, such as the driving strengths and the detuning, these distinct transition pathways can be made to destructively interfere with each other. Mathematically, this perfect destructive interference is explicitly manifested by the condition $|B|^2 = 0$. As defined in Eq.~(\ref{eqb}), since $|B|^2$ is composed of the sum of two independent squared terms, the condition $|B|^2 = 0$ strictly requires both terms to vanish simultaneously. By setting the second term to zero, we can analytically extract the optimal detuning condition:
	\begin{align}
		\Delta_0 = \frac{2\lambda G (\kappa_a+\gamma_q)}{\kappa_a+3\gamma_q}.
		\label{d0}
	\end{align}
	Subsequently, the corresponding optimal driving ratio $\lambda$ can be determined by substituting $\Delta_0$ back into the equation of the first term being zero. Solving this equation directly yields the exact analytical expression for the optimal driving ratio:
	\begin{align}
		\lambda_{\text{opt}} =\sqrt{ \frac{G^2 - \frac{\gamma_q(\kappa_a+\gamma_q)}{4}}{G^2 \left[ \frac{16\gamma_q(\kappa_a+\gamma_q)}{(\kappa_a+3\gamma_q)^2} - 1 \right]} }.
	\end{align}
	Under these exact analytical solutions, the transition probability to the two-photon state is completely suppressed, meaning $|C_{2,g}|^2 = 0$. Therefore, the equal-time second-order correlation function $g^{(2)}(0)$ ultimately approaches zero. Furthermore, to achieve UPB, it is not strictly necessary to continuously vary the driving ratio $\lambda$ with the coupling strength $G$. For a suitably chosen fixed $\lambda$, one can still effectively suppress the two-photon transition ($|C_{2,g}|^2 \to 0$) and achieve strong blockade by finely tuning the detuning $\Delta_0$ around its analytically predicted optimal value.
	
	However, this parameter flexibility is entirely lost in the single-drive configuration ($\lambda = 0$). By substituting $\lambda = 0$ into the quantum interference condition $|B|^2 = 0$, the vanishing of the second term immediately dictates a zero detuning, $\Delta_0 = 0$. Subsequently, substituting both $\lambda = 0$ and $\Delta_0 = 0$ into the first term being zero yields a remarkably rigid constraint on the system's coupling strength itself:
	\begin{align}
		G = \frac{\sqrt{\gamma_q(\kappa_a+\gamma_q)}}{2}.
	\end{align}
	This derivation reveals that the single-drive UPB can only be observed at a isolated parameter point. Consequently, the single-drive scheme is incapable of realizing photon blockade over a broad range of coupling strengths. In stark contrast, the introduction of dual drive liberates the system from this strict requirement, enabling robust UPB across a much wider range of coupling strengths. 
	
\begin{figure}[tb]
		\centering
		\includegraphics[width=0.6\linewidth]{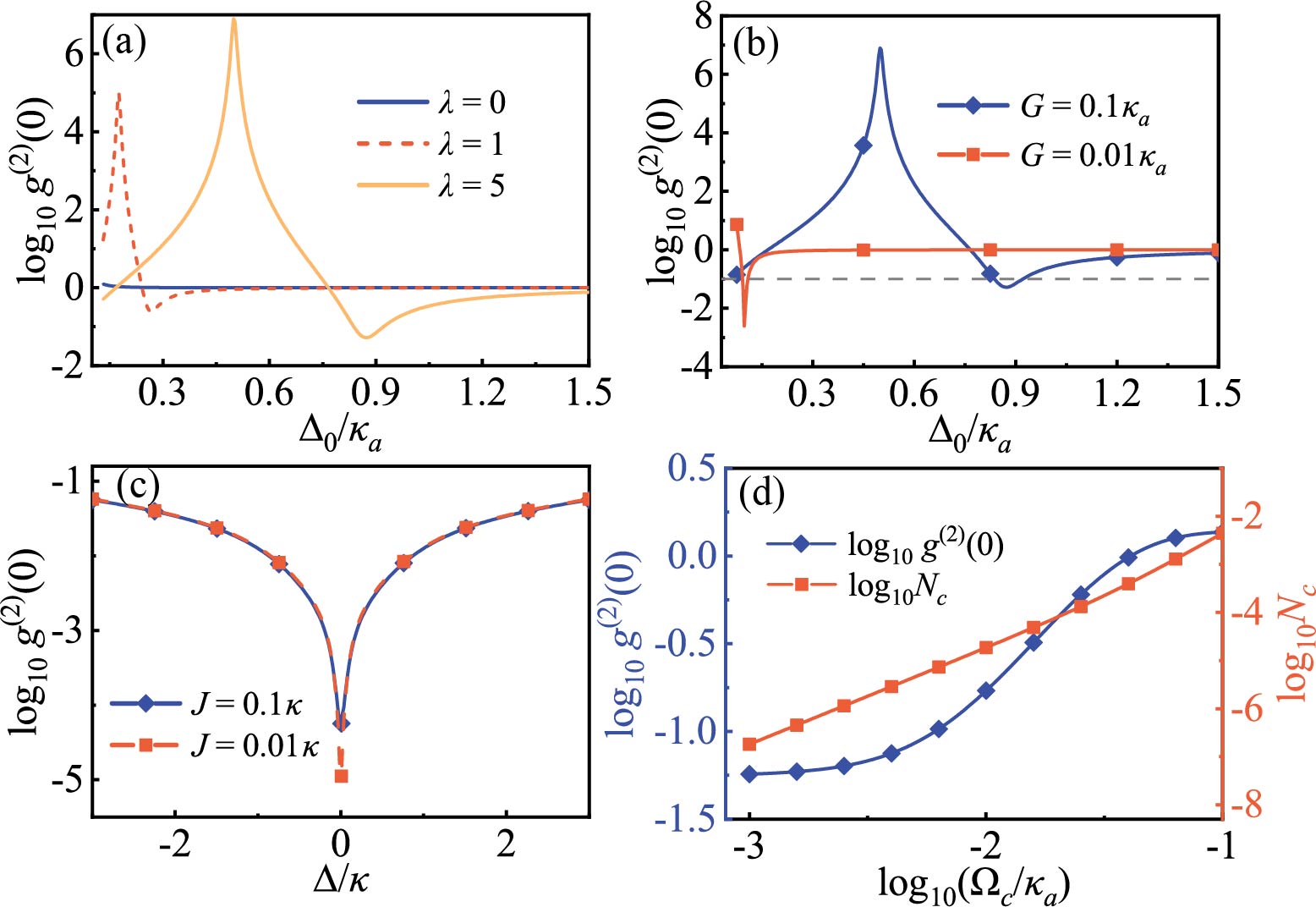} 
		\caption{Performance of the UPB in the extreme weak-coupling regime. (a) The equal-time second-order correlation function $g^{(2)}(0)$ as a function of the normalized detuning $\Delta_0/\kappa_a$ for different driving ratios $\lambda$ with $G=0.1\kappa_a$. (b) $g^{(2)}(0)$ versus $\Delta_0/\kappa_a$ for varying coupling strengths with the driving ratio $\lambda=5$. The horizontal dashed gray line at $\log_{10}g^{(2)}(0) = -1$ defines the threshold for the effective blockade window. (c) $g^{(2)}(0)$ as a function of the normalized detuning $\Delta/\kappa$ for our proposed novel mechanism in the extremely weak-coupling regime ($J=0.1\kappa$ and $J=0.01\kappa$). (d) $g^{(2)}(0)$ (left blue axis) and the intracavity mean photon number $N_c$ (right red axis) as functions of the normalized driving strength $\Omega_c/\kappa_a$ for the UPB with $G=0.1\kappa_a$ and $\lambda=5$. 
			Unless otherwise specified, the system parameters are set to $\gamma_q=0.01\kappa_a$ and $\Omega_c=0.01G$.}
		\label{fig3}
	\end{figure}

	As shown in Fig.~\ref{fig3}(a), we plot the equal-time second-order correlation function $g^{(2)}(0)$ as a function of the normalized detuning $\Delta_0/\kappa_a$ with different ratios $\lambda$. As explicitly depicted, the single-drive case ($\lambda=0$) completely fails to achieve blockade. However, by introducing the dual-drive configuration and tuning the ratio to $\lambda=5$, strong photon blockade is achieved, confirming that the dual-drive scheme effectively extends the UPB into the extremely weak-coupling regime.
	
	Nevertheless, the UPB mechanism exhibits a high sensitivity to system parameters. To quantitatively evaluate this sensitivity, we define the effective blockade window as the continuous detuning range where $g^{(2)}(0) < 0.1$ [indicated by the horizontal dashed gray line at $\log_{10}g^{(2)}(0) = -1$ in Fig.~\ref{fig3}(b)]. As illustrated, while the system sustains an effective blockade window spanning a range of approximately $0.1\kappa_a$ for $G=0.1\kappa_a$, this window drastically shrinks to a narrow range of only $0.01\kappa_a$ when the coupling strength is reduced to $G=0.01\kappa_a$. This stark contrast clearly demonstrates that achieving UPB in the extremely weak-coupling regime imposes stringent requirements on the precision of the frequency detuning.
	
	In striking contrast, our proposed mechanism exhibits exceptional robustness against parameter variations, as evidently demonstrated in Fig.~\ref{fig3}(c). Here, we plot $g^{(2)}(0)$ versus the normalized detuning $\Delta/\kappa$ for our scheme under extremely weak coupling conditions ($J=0.1\kappa$ and $J=0.01\kappa$). Remarkably, the blockade profiles for different coupling strengths nearly overlap, maintaining a broad blockade window of approximately $6\kappa$ (roughly from $\Delta = -3\kappa$ to $3\kappa$). This confirms that our scheme maintains a robust photon blockade from the strong-coupling regime all the way down to the ultra-weak-coupling regime, thereby overcoming the extreme parameter sensitivity that limits the traditional UPB.
	
	Furthermore, the UPB is also limited by a severe trade-off between single-photon purity and source brightness. As depicted in Fig.~\ref{fig3}(d), as the driving strength $\Omega_c$ increases, the mean photon number $N_c$ predictably rises; however, the correlation function $g^{(2)}(0)$ simultaneously degrades. More importantly, unlike the CPB mechanism which benefits from single-photon resonant excitation, the UPB lacks such resonant enhancement. Consequently, its mean photon number $N_c$ remains much smaller, directly resulting in a lower source brightness. To maintain a strong photon blockade, the maximum tolerable driving strength must scale proportionally with the coupling strength $G$. Therefore, a weaker coupling necessitates a correspondingly reduced driving field. To satisfy this weak-driving condition, the driving strength is set to $\Omega_c = 0.01G$. For the remaining parameters about the UPB calculations in Fig.~3 of the main text, we set $\lambda = 5$ and $\gamma_q = 0.01\kappa_a$, with the optimal detuning continuously adjusted via Eq.~(\ref{d0}) as the $\kappa_a/G$ varies.
	
	\section{\label{SE3} Physical Realization of the Three-Body Interaction}
	
	In this section, we present the detailed circuit architecture and theoretical derivation. We propose a scheme to realize the target three-body interaction between two transmon qubits and a coplanar waveguide (CPW) resonator via a flux-tunable Josephson coupler. Specifically, the coupler is directly connected to the two transmon qubits, while one of its junctions is connected in series with a localized constricted segment of the CPW center conductor. In the following subsections, we will systematically formulate the Hamiltonians for the resonator, the qubits, and the coupling circuit to demonstrate how the pure three-body interaction is rigorously generated and the residual two-body terms are effectively suppressed.
	
	\subsection{Quantization of the CPW Resonator}
	
	The CPW resonator~\cite{RevModPhys.93.025005,PhysRevLett.124.193604} of length $l$ is treated as a linear, dispersion-free one-dimensional transmission line governed by the continuous Hamiltonian
	\begin{align}
		H_{\text{cpw}} = \int_{0}^{l} dx \left\{ \frac{1}{2c_0} Q(x,t)^2 + \frac{1}{2l_0} [\partial_x \Phi(x,t)]^2 \right\},
		\label{eqcp}
	\end{align}
	where $c_0$ and $l_0$ denote the capacitance and inductance per unit length, respectively, $Q(x,t)$ is the charge density field, and $\Phi(x,t)$ is the continuous flux field. Using Hamilton's equations together with Eq.~(\ref{eqcp}), the resulting wave equation for the flux propagation is
	\begin{align}
		\frac{1}{l_0 c_0} \frac{\partial^2 \Phi(x,t)}{\partial x^2} - \frac{\partial^2 \Phi(x,t)}{\partial t^2} = 0.\label{eqbo}
	\end{align}
	The solution to Eq.~(\ref{eqbo}) are expanded in terms of the discrete spatial normal modes as
	\begin{align}
		\Phi(x,t) = \sum_{n=1}^{\infty} \Phi_n(t) u_n(x), \quad Q(x,t) = \frac{1}{l} \sum_{n=1}^{\infty} Q_n(t) u_n(x),
	\end{align}
	For a open-ended $\lambda/2$ resonator, the current vanishes at both ends, imposing the boundary conditions $\partial_x \Phi(0,t) = \partial_x \Phi(l,t) = 0$. Consequently, the normalized spatial mode profile is given by 
	\begin{equation}
		u_n(x) = \sqrt{2} \cos(k_n x),
	\end{equation}
	with the discrete wave vector $k_n = n\pi / l$, satisfying the orthogonality condition $\frac{1}{l} \int_0^l dx u_n(x) u_m(x) = \delta_{nm}$.
	
	Substituting the normal mode expansions into the continuous Hamiltonian and performing the spatial integration, the Hamiltonian in Eq.~(\ref{eqcp}) can be expressed as
	\begin{align}
		H_{\text{cpw}} = \sum_{n=1}^{\infty} \left( \frac{Q_n^2}{2C} + \frac{C\omega_n^2\Phi_n^2}{2} \right),
	\end{align}
	where $C = c_0 l$ is the total capacitance, and $\omega_n=k_n/\sqrt{c_0v_0}$.
	
	$L_n$ is the effective inductance of the $n$-th mode with the corresponding resonance frequency $\omega_n = 1/\sqrt{L_n C}$.
	In the canonical quantization procedure, the conjugate variables are promoted to quantum operators satisfying $[\hat{\Phi}_n, \hat{Q}_m] = i\hbar \delta_{nm}$. Introducing the bosonic creation $\hat{a}_n^\dagger$ and annihilation $\hat{a}_n$ operators, the flux and charge operators are written as 
	\begin{align}
		\hat{\Phi}_n = \sqrt{\hbar Z_n / 2} (\hat{a}_n^\dagger + \hat{a}_n), \\
		\hat{Q}_n = \sqrt{\hbar  / 2Z_n} (\hat{a}_n^\dagger - \hat{a}_n),
	\end{align}
	where $Z_n = \sqrt{L_n / C}$ is the characteristic impedance of $n$ with $1/L_n=C\omega_n^2$. Substituting this expression, we obtain the final result
	\begin{equation}
		\hat{H}_\text{cpw} = \sum_{n=1}^{\infty}\hbar \omega_n \hat{a}^\dagger \hat{a}
	\end{equation}
	
	In this work, we consider a three-body resonance among the fundamental mode ($n=1$) and two qubits, while all higher-order modes are far detuned. The system can therefore be safely truncated to the fundamental mode with resonance frequency $\omega_a=\omega_1$ . Retaining only the $n=1$ terms, the free Hamiltonian of resonator reduces to $\hat{H}_\text{cpw} = \hbar \omega_a \hat{a}^\dagger \hat{a}$, and the corresponding spatial flux operator simplifies to
	\begin{align}
		\hat{\Phi}(x) = \sqrt{\frac{\hbar Z_1}{2}} (\hat{a}_n^\dagger + \hat{a}_n) \sqrt{2} \cos\left(\frac{\pi x}{l}\right).
	\end{align}
	
\begin{figure}[tb]    
		\centering    
		\includegraphics[width=1\textwidth]{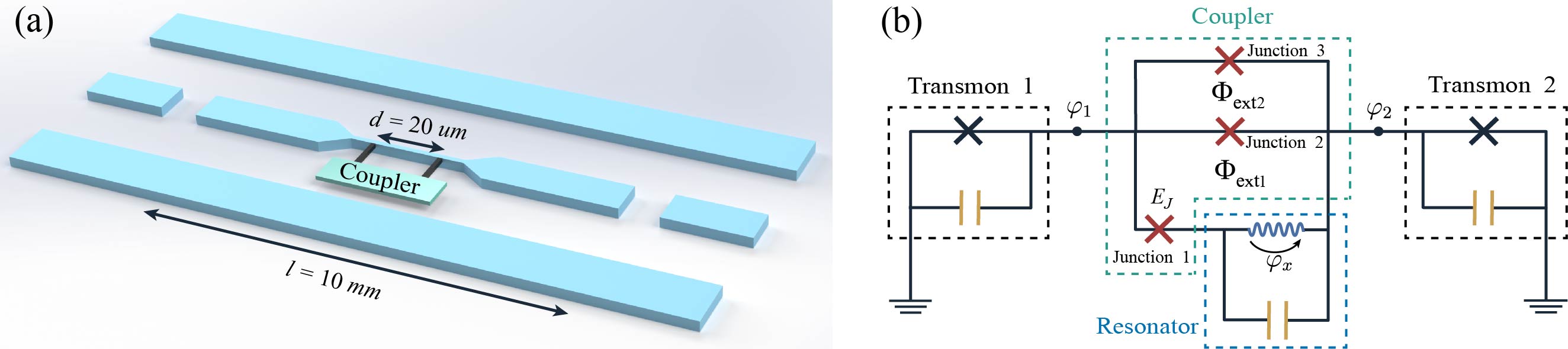}    
		\caption{Schematic of the hardware architecture for the three-body interaction. (a) 3D physical layout of the device. A coplanar waveguide (CPW) resonator of total length $l = 10\text{ mm}$ features a localized geometrical constriction at its center. The Josephson coupler is connected across a distance of $d = 20\ \mu\text{m}$ within this highly inductive region to maximize the local flux gradient. (b) The corresponding circuit diagram. Two transmon qubits, denoted by their node phases $\varphi_1$ and $\varphi_2$, are directly connected to the terminals of the coupler. The coupler consists of three parallel branches with Josephson junctions (Junctions 1-3) to simultaneously mediate the target three-body interaction and actively cancel the residual two-body interaction. The two superconducting loops are threaded by external magnetic fluxes $\Phi_{\text{ext1}}$ and $\Phi_{\text{ext2}}$ for in situ tunability. Notably, Junction 1 is connected in series with the constricted CPW segment, which is characterized by a localized phase difference $\varphi_x$. }    
		\label{FIG4}
	\end{figure}

	To introduce the coupling, the Josephson coupler, typically realized as a tunable SQUID loop, is embedded into the center conductor of the CPW so that the segment of length $d$ of the center conductor is connected into the coupler’s circuit, as schematically shown in Fig.~\ref{FIG4}(a). The coupler is connected over the segment from $x=(l+d)/2$ to $x=(l-d)/2$. The phase difference across this localized segment of length $d$ is determined by the integration of the local flux gradient:
	\begin{align}
		\hat{\varphi}_x = \frac{2\pi}{\Phi_0} \int_{(l-d)/2}^{(l+d)/2} \partial_{x} \hat{\Phi}(x) dx \approx \frac{2\pi d}{\Phi_0} \partial_x \hat{\Phi}(x)\Big|_{x=l/2},
	\end{align}
	where $\Phi_0 = h/2e$ is the magnetic flux quantum. Since $d\ll l$, the flux gradient $\partial_x \hat{\Phi}(x)$ is essentially constant over the segment, so the integral simplifies to the product of $d$ and the gradient evaluated at the midpoint. For a uniform center conductor, evaluating the spatial derivative of the fundamental mode profile at $x = l/2$ yields $\partial_x u_1(l/2) = \sqrt{2}\pi/l$.
	
	The inclusion of a geometrical constriction within this localized segment increases the local inductance per unit length, thereby steepening the spatial gradient of the flux field. This constriction increases the gradient by a factor of several times~\cite{ PhysRevA.80.032109, PhysRevB.78.180502, PhysRevApplied.16.044045,Gimeno2020}, and the phase difference is scaled by a phenomenological constriction enhancement factor $\eta$. The effective phase difference operator across the constricted segment is ultimately expressed as
	\begin{align}
		\hat{\varphi}_x = \eta \frac{2\sqrt{2}\pi^2 d}{\Phi_0 l} \sqrt{\frac{\hbar Z_1}{2}} (\hat{a}^\dagger + \hat{a}). \label{eq:phix}
	\end{align}
	
	\subsection{The Transmon Qubits}
	
	Each transmon is realized by a Josephson junction with Josephson energy $E_{Ji}$ shunted by a capacitance $C_i$ (with the corresponding charging energy $E_{Ci}=e^2/(2C_i)$). The bare Hamiltonian for transmon $i$ ($i=1,2$) is given by
	\begin{equation}
		\hat{H}_{qi} = 4E_{Ci}\,\hat{n}_i^2 - E_{Ji}\cos\hat{\varphi}_i,
	\end{equation}
	where $\hat{\varphi}_i$ is the quantum phase difference across the junction and $\hat{n}_i$ is the conjugate Cooper-pair number operator. They satisfy the canonical commutation relation $[\hat{\varphi}_i, \hat{n}_j] = i\delta_{ij}$.
	
	Introducing the bosonic creation $\hat{b}_i^\dagger$ and annihilation $\hat{b}_i$ operators~\cite{RevModPhys.73.357,doi:10.1126/science.239.4843.992}, the phase and charge operators are quantized as
	\begin{align}
		\hat{\varphi}_i &= \left( \frac{2E_{Ci}}{E_{Ji}} \right)^{1/4} (\hat{b}_i^\dagger + \hat{b}_i), \\
		\hat{n}_i &= i \left( \frac{E_{Ji}}{32E_{Ci}} \right)^{1/4} (\hat{b}_i^\dagger - \hat{b}_i).
	\end{align}
	
	Substituting these expressions into the bare Hamiltonian and expanding the cosine potential to the fourth order in $\hat{\varphi}_i$ (due to $E_{Ji} \gg E_{Ci}$ in the transmon regime), we obtain the standard Duffing oscillator Hamiltonian. After applying the rotating-wave approximation (RWA) to eliminate the rapidly oscillating, non-number-conserving terms, the Hamiltonian reduces to
	\begin{equation}
		\hat{H}_{qi} \approx \hbar\omega_i\,\hat{b}_i^\dagger\hat{b}_i - \frac{E_{Ci}}{2}\,\hat{b}_i^\dagger\hat{b}_i^\dagger\hat{b}_i\hat{b}_i,
	\end{equation}
	with the fundamental transition frequency defined as $\hbar\omega_i = \sqrt{8E_{Ci}E_{Ji}} - E_{Ci}$. 
	
	It is noteworthy that due to the system's inherent anharmonicity (quantified by $-E_{Ci}$), multi-photon excitations to higher energy levels are effectively prevented. We can thus safely restrict the system dynamics to the computational subspace formed by the lowest two energy levels: the ground state $|g\rangle$ and the excited state $|e\rangle$. Under this two-level approximation, the bosonic annihilation and creation operators can be directly mapped to the Pauli lowering and raising operators, respectively, i.e.,
	\begin{equation}
		\hat{b}_i \rightarrow \hat{\sigma}_i^-, \quad \hat{b}_i^\dagger \rightarrow \hat{\sigma}_i^+.
	\end{equation}
	Correspondingly, the transmon phase operator projected into this qubit subspace takes the form $\hat{\varphi}_i \approx (2E_{Ci}/E_{Ji})^{1/4} (\hat{\sigma}_i^+ + \hat{\sigma}_i^-)$. As a result, the free Hamiltonian of each transmon reduces to an effective two-level form:
	\begin{equation}
		\hat{H}_{qi} = \hbar\omega_i \hat{\sigma}_i^+ \hat{\sigma}_i^-,
	\end{equation}
	where $\hat{\sigma}_i^+ \hat{\sigma}_i^- = |e\rangle\langle e|$ represents the excitation number operator in the two-level subspace. The total free Hamiltonian of the uncoupled two-transmon system is ultimately written as $\hat{H}_q = \hat{H}_{q1} + \hat{H}_{q2}$. 
	
	To experimentally realize the target three-body resonance condition derived in the previous section, the circuit parameters must be carefully engineered. Specifically, we set the Josephson energies of the two transmons as $E_{J1}/h = 45\text{ GHz}$ and $E_{J2}/h = 10\text{ GHz}$, with their corresponding charging energies configured to $E_{C1}/h = 0.3\text{ GHz}$ and $E_{C2}/h = 0.2\text{ GHz}$. These specific choices yield the effective qubit transition frequencies of $\omega_1/2\pi \approx 10\text{ GHz}$ and $\omega_2/2\pi \approx 4\text{ GHz}$. By designing the WW resonator to have a fundamental frequency of $\omega_a/2\pi = 6\text{ GHz}$, the strict energy conservation condition $\omega_1 = \omega_a + \omega_2$ is satisfied. This precise frequency matching activates the desired three-body interaction $\hat{H}_{\text{thr}}$ while safely suppressing other non-resonant multi-photon processes.
	\subsection{Generation of Three-Body Interaction}
	
	As schematically illustrated in Fig.~\ref{FIG4}(b), the two transmon qubits are directly connected to the terminals of the coupler. The coupler consists of three parallel branches tailored to simultaneously mediate the target interaction and cancel unwanted interaction. Specifically, Junction 1 is utilized to generate the desired three-body interaction, whereas Junction 2 and Junction 3 (with scaled energies $\alpha E_J$ and $\beta E_J$, respectively) are designed to completely eliminate the residual two-body interactions. To achieve the three-body coupling, Junction 1 (with energy $E_J$) is connected in series with the localized constricted segment of the CPW center conductor, thereby physically integrating the localized phase difference $\hat{\varphi}_x$ of the resonator into the circuit dynamics.
	
	To derive the interaction, we first determine the phase difference across each Josephson junction using the phase constraints imposed by flux quantization. Let the node phases of the two transmons be $\hat{\varphi}_1$ and $\hat{\varphi}_2$. Since Junction 2 directly connects the two transmon qubits, the phase difference across it is $\hat{\varphi}_{J2} = \hat{\varphi}_1 - \hat{\varphi}_2$. For Junction 1, which is connected in series with the CPW segment, its phase difference is affected by both the magnetic flux $\Phi_{\text{ext1}}$ threading the first loop and the local phase difference of the CPW, yielding $\hat{\varphi}_{J1} = \hat{\varphi}_1 - \hat{\varphi}_2 - \phi_{\text{ext1}} - \hat{\varphi}_x$. Similarly, the magnetic flux $\Phi_{\text{ext2}}$ through the second loop influences the phase difference across Junction 3 as $\hat{\varphi}_{J3} = \hat{\varphi}_1 - \hat{\varphi}_2 + \phi_{\text{ext2}}$.
	
	Consequently, the total Hamiltonian of the coupled system can be constructed as the sum of the free Hamiltonians and the Josephson potentials of the three Josephson junctions:
	\begin{align}
		\hat{H} &= \hat{H}_{q1} + \hat{H}_{q2} + \hat{H}_{\text{cpw}} 
		- E_J \cos(\hat{\varphi}_1 - \hat{\varphi}_2 - \phi_{\text{ext1}} - \hat{\varphi}_x) - \alpha E_J \cos(\hat{\varphi}_1 - \hat{\varphi}_2)
		- \beta E_J \cos(\hat{\varphi}_1 - \hat{\varphi}_2 + \phi_{\text{ext2}}).
	\end{align}
	
	The physical origin of the three-body interaction lies in the potential of Junction 1, which simultaneously incorporates the degrees of freedom of both transmons and the resonator. Since only a small segment of the resonator is connected to the circuit ($d \ll l$), the local phase fluctuations are sufficiently small ($\hat{\varphi}_x \ll 1$). We can thus expand the potential of Junction 1 via Taylor series to the first order in $\hat{\varphi}_x$:
	\begin{align}
		\hat{H}_{J1} &= -E_J \cos(\hat{\varphi}_1 - \hat{\varphi}_2 - \phi_{\text{ext1}}) - E_J \hat{\varphi}_x \sin(\hat{\varphi}_1 - \hat{\varphi}_2 - \phi_{\text{ext1}}).
	\end{align}
	This expansion clearly separates the potential into two distinct parts.  The first term is a static $0$-th order component that induces residual two-body crosstalk between the transmons, and we will rigorously group this term with the potentials of junction 2 and 3 to completely eliminate it in the subsequent subsection. The second term, proportional to $\hat{\varphi}_x$, governs the dynamic energy exchange and is isolated as the interaction Hamiltonian $\hat{H}_{\text{int}}$:
	\begin{align}
		\hat{H}_{\text{int}} &= -E_J \hat{\varphi}_x (\hat{\varphi}_1 - \hat{\varphi}_2) \cos\phi_{\text{ext1}} + E_J \hat{\varphi}_x \left( 1 + \hat{\varphi}_1\hat{\varphi}_2 - \frac{\hat{\varphi}_1^2}{2} - \frac{\hat{\varphi}_2^2}{2} \right) \sin\phi_{\text{ext1}},
		\label{eq:Hint}
	\end{align}
	where we have further expanded the transmon phase difference up to the second order using small-angle approximations.
	
	Substituting the resonator phase operator $\hat{\varphi}_x$ using the bosonic operators from Eq.~(\ref{eq:phix}) and expressing the transmon phase operators in terms of Pauli spin operators $\hat{\varphi}_j \approx (2E_{Cj}/E_{Jj})^{1/4} (\hat{\sigma}_j^+ + \hat{\sigma}_j^-)$ into Eq.~(\ref{eq:Hint}), the dynamic interaction Hamiltonian is explicitly expanded as:
	\begin{align}
		\hat{H}_{\text{int}} = - \hbar(\hat{a}^\dagger + \hat{a}) \left[ g_1 (\hat{\sigma}_1^+ + \hat{\sigma}_1^-) - g_2 (\hat{\sigma}_2^+ + \hat{\sigma}_2^-) \right]  + \hbar J  (\hat{a}^\dagger + \hat{a})(\hat{\sigma}_1^+ + \hat{\sigma}_1^-)(\hat{\sigma}_2^+ + \hat{\sigma}_2^-),
	\end{align}
	
\begin{figure}[tb]    
		\centering    
		\includegraphics[width=0.6\textwidth]{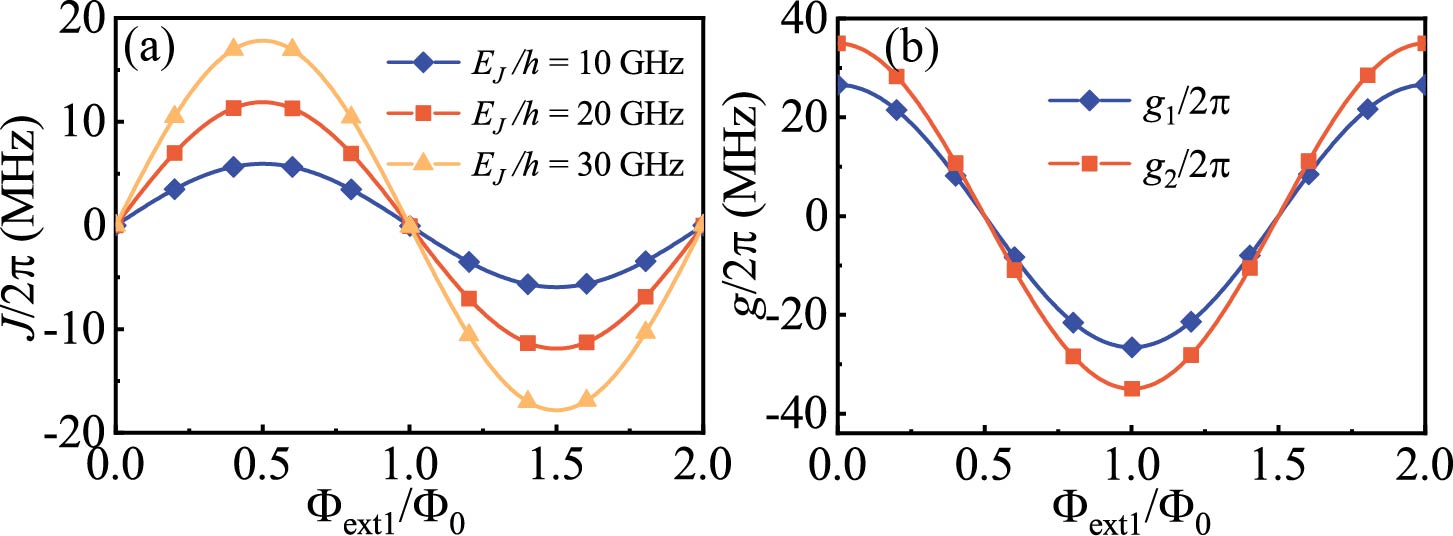}    
		\caption{(a) The three-body coupling strength $J/2\pi$ as a function of the external magnetic flux $\Phi_{\text{ext1}}/\Phi_0$ for different Josephson energies $E_J$. (b) The two-body coupling strengths $g_1/2\pi$ and $g_2/2\pi$ as a function of the external magnetic flux $\Phi_{\text{ext1}}/\Phi_0$ with $E_J/h = 20\text{GHz}$. Here, we choose constriction enhancement factor $\eta=5$. }    
		\label{FIG5}
	\end{figure}

	As illustrated in Fig.~\ref{FIG5}, we plot the coupling strengths as a function of the normalized external magnetic flux $\Phi_{\text{ext1}}/\Phi_0$. Specifically, Fig.~\ref{FIG5}(a) shows the three-body coupling strength $J/2\pi$ for different values of the coupler's Josephson energy $E_J$. It is clearly observed that with $E_J/h = 20\text{ GHz}$, the three-body coupling strength can readily reach over $10\text{ MHz}$ at the optimal bias point $\Phi_{\text{ext1}}/\Phi_0 = 0.5$. Simultaneously, as depicted in Fig.~\ref{FIG5}(b), when $J$ attains its maximum value, the residual two-body coupling strengths $g_1$ and $g_2$ exactly vanish to zero, which is a direct mathematical consequence of their cosine dependence. 
	
	Furthermore, even if the external flux fluctuates and the system operates away from this sweet spot, the two-body couplings can still be safely neglected. Although $g_1$ and $g_2$ might reach their maximal values of several tens of MHz under non-ideal bias, the frequency detunings between the resonator and the two qubits are defined as $\Delta_1/2\pi = |\omega_{q1} - \omega_c|/2\pi = 4\text{ GHz}$ and $\Delta_2/2\pi = |\omega_c - \omega_{q2}|/2\pi = 2\text{ GHz}$, respectively. The frequency detunings are orders of magnitude larger than the maximum possible two-body coupling strengths ($g_j \ll \Delta_j$). Consequently, by applying the rotating-wave approximation (RWA) to eliminate these non-energy-conserving terms under the resonance condition $\omega_1=\omega_a+\omega_2$, we successfully obtain the target three-body interaction
	\begin{align}
		\hat{H}_{\text{thr}} = J \left( \hat{a}^\dagger \hat{\sigma}_1^- \hat{\sigma}_2^+ + \hat{a} \hat{\sigma}_1^+ \hat{\sigma}_2^- \right).
	\end{align}

	\subsection{Suppression of Residual Two-Body Interactions}
	
	As emphasized in the previous derivation, setting $\phi_{\text{ext1}} = \pi/2$ successfully achieve the optimal three-body interaction, but it leaves behind the  component of Junction 1: $-E_J \cos(\hat{\varphi}_1 - \hat{\varphi}_2 - \pi/2) = -E_J \sin(\hat{\varphi}_1 - \hat{\varphi}_2)$. To fully describe the unwanted direct two-body interactions between the qubits, this residual terms must be systematically grouped with the potentials of Junction 2 and Junction 3. The total residual two-body Hamiltonian is thus assembled as
	\begin{align}
		\hat{H}_{\text{oth}} ={- E_J \sin(\hat{\varphi}_1 - \hat{\varphi}_2)}{-\alpha E_J \cos(\hat{\varphi}_1 - \hat{\varphi}_2)} {- \beta E_J \cos(\hat{\varphi}_1 - \hat{\varphi}_2 + \phi_{\text{ext2}})}.
	\end{align}
	Expanding the third term and regrouping the sine and cosine components, the expression is reorganized as
	\begin{align}
		\hat{H}_{\text{oth}} = -E_J \left( \alpha + \beta \cos\phi_{\text{ext2}} \right) \cos(\hat{\varphi}_1 - \hat{\varphi}_2) - E_J \left( 1 - \beta \sin\phi_{\text{ext2}} \right) \sin(\hat{\varphi}_1 - \hat{\varphi}_2).
	\end{align}
	To strictly ensure the purity of the three-body interaction, these static two-body couplings must be perfectly suppressed. This mathematically requires the coefficients of both the cosine and sine terms to vanish simultaneously:
	\begin{align}
		\cos\phi_{\text{ext2}} = -\frac{\alpha}{\beta}, 
		\sin\phi_{\text{ext2}} = \frac{1}{\beta}.
	\end{align}
	By solving these two exact constraints, we discover that the junction asymmetry parameters must inherently satisfy the geometric relation $\beta = \sqrt{\alpha^2 + 1}$. When the circuit is fabricated to meet this specific condition and the second loop is biased at the designated flux $\phi_{\text{ext2}} = \arccos(-\alpha/\beta)$, the residual two-body interactions $\hat{H}_{\text{oth}}$ are completely intrinsically cancelled. However, it is worth noting that our system does not strictly rely on perfectly satisfying this ideal condition. Since the frequency detuning between the two transmon qubits is profoundly large ($\Delta_{12}/2\pi = |\omega_1 - \omega_2|/2\pi = 6\text{ GHz}$), any residual direct qubit-qubit exchange interaction caused by realistic fabrication imperfections or flux bias deviations remains highly off-resonant. Consequently, this unwanted two-body interaction is naturally suppressed by the large detuning and can still be safely neglected.
	
	\bibliography{SMReference}
	\bibliographystyle{apsrevlong}